\global\def\draftcontrol{0}

   \def\versionno{ n=2 thermo -- draft   }

\catcode`\@=11

\expandafter\ifx\csname draftcontrol\endcsname\relax\global\def\draftcontrol{0}
\fi

{\count255=\time\divide\count255 by 60
\xdef\hourmin{\number\count255}
\multiply\count255 by-60\advance\count255 by\time
\xdef\hourmin{\hourmin:\ifnum\count255<10 0\fi\the\count255}}
\def\draftdate{\number\month/\number\day/\number\year\ \ \ \hourmin }

\newcommand\makepapertitle{\par
  \begingroup
    \renewcommand\thefootnote{\@fnsymbol\c@footnote}%
    \def\@makefnmark{\rlap{\@textsuperscript{\normalfont\@thefnmark}}}%
    \long\def\@makefntext##1{\parindent 1em\noindent
            \hb@xt@1.8em{%
                \hss\@textsuperscript{\normalfont\@thefnmark}}##1}%
     \newpage
     \global\@topnum\z@   
     \@makepapertitle
     \thispagestyle{empty}\@thanks
  \endgroup
  \setcounter{footnote}{0}%
  \global\let\thanks\relax
  \global\let\makepapertitle\relax
  \global\let\@makepapertitle\relax
  \global\let\@thanks\@empty
  \global\let\@author\@empty
  \global\let\@date\@empty
  \global\let\@title\@empty
  \global\let\title\relax
  \global\let\author\relax
  \global\let\date\relax
  \global\let\and\relax
  \def\version{\let\version\@version\@gobble}
}
\def\@makepapertitle{%
  \newpage
   \ifnum\draftcontrol=1 {}
   \version\versionno
   \vskip 3em%
   \else
   \hfill\hbox to 3cm {\parbox{4cm}{\@pubnum}\hss}%
   \vskip 3em%
   \fi
   \begin{center}%
   \let \footnote \thanks
     {\LARGE {\@title}}%
     \vskip 1.5em%
     {\normalsize
       \lineskip .5em%
       \begin{tabular}[t]{c}%
         \@author
       \end{tabular}\par}%
     \vskip 1.5em%
     {\@bstract}%
     \end{center}%
     \vskip 1.5em
     \@date%
   \par
}

\gdef\@pubnum{}
\def\pubnum#1{%
  \gdef\@pubnum{#1}}

\gdef\@bstract{}
\def\Abstract#1{%
  \gdef\@bstract{%
   \parbox{\textwidth-0pc}{%
   \centerline{\bf Abstract}\penalty1000%
\kern.2cm%
\noindent
\renewcommand\baselinestretch{1.0}%
{#1}}}
}

\def\ps@paper{\let\@mkboth\@gobbletwo%
     \ifnum\draftcontrol=1
	\def\@oddfoot{\hbox to \textwidth{\tiny \versionno \hfil\tiny\draftdate}%
	\hskip -\textwidth \hbox to \textwidth{\hfil\rm\thepage\hfil}}%
     \else\def\@oddfoot{\hbox to \textwidth{\hfil\rm\thepage\hfil}}
     \fi
     \let\@evenfoot\@oddfoot
}

\def\body{\clearpage
          \pagestyle{paper}
	}

\def\@version#1{\ifnum\draftcontrol=1
\typeout{}\typeout{#1}\typeout{}
\vskip3mm\centerline{\hbox{\fbox{\normalsize{\tt DRAFT -- #1 -- }
                   {\draftdate}}}}\vskip3mm
\fi}
\let\version\@version
\long\def\eqlabel#1{\ifnum\draftcontrol=1
                    \tag@false  
                    \tag*{(\theequation) \hbox to -0.2cm{\hspace{0cm}\small{#1}\hss}}
                    \refstepcounter{equation}
                    \edef\@currentlabel{\theequation}
                    \ltx@label{#1}          
                    \else
                    \label{#1}
                    \fi
                    }
\let\st@bibitem\@bibitem
\let\st@lbibitem\@lbibitem
\ifnum\draftcontrol=1
  \def\@bibitem#1{%
    \st@bibitem{#1}\a@@label{#1}\ignorespaces}
  \def\@lbibitem[#1]#2{%
    \st@lbibitem[#1]{#2}\a@@label{#2}\ignorespaces}
  \def\a@@label#1{%
    \gdef\a@lab{\smash{\normalfont\small#1}}
    \ifvmode
      \if@inlabel
        \global\setbox\@labels\hbox{%
          \llap{\a@lab\let\a@lab\relax
                \kern\@totalleftmargin\kern\marginparsep}%
          \box\@labels}%
      \fi
    \fi}
\fi

\documentclass[12pt,letterpaper]{article}

\usepackage{amsmath,amssymb,array,calc,rotating,epsfig,psfrag}
\usepackage[nosort]{cite}

\ifnum\draftcontrol=1
\fi

\renewcommand\baselinestretch{1.25}
\renewcommand\section{\@startsection {section}{1}{\z@}%
                                   {-3.5ex \@plus -1ex \@minus -.2ex}%
                                   {2.3ex \@plus.2ex}%
                                   {\normalfont\large\bfseries}}
\renewcommand\subsection{\@startsection{subsection}{2}{\z@}%
                                   {-3.25ex\@plus -1ex \@minus -.2ex}%
                                   {1.5ex \@plus .2ex}%
                                   {\normalfont\normalsize\bfseries}}
\renewcommand\subsubsection{\@startsection{subsubsection}{3}{\z@}%
                                   {-3.25ex\@plus -1ex \@minus -.2ex}%
                                   {1.5ex \@plus .2ex}%
                                   {\normalfont\normalsize\it}}
\renewcommand\paragraph{\@startsection{paragraph}{4}{\z@}%
                                   {-3.25ex\@plus -1ex \@minus -.2ex}%
                                   {1.5ex \@plus .2ex}%
                                   {\normalfont\normalsize\bf}}

\numberwithin{equation}{section}

\def\ie{{\it i.e.}}

\def\revise#1       {\raisebox{-0em}{\rule{3pt}{1em}}%
                     \marginpar{\raisebox{.5em}{\vrule width3pt\
                     \vrule width0pt height 0pt depth0.5em
                     \hbox to 0cm{\hspace{0cm}{%
                     \parbox[t]{4em}{\raggedright\footnotesize{#1}}}\hss}}}}

\newcommand\nxt[1]  {\\\fnxt#1}

\def\cala         {{\cal A}}

\def\calb         {{\cal B}}

\def\calf         {{\cal F}}

\def\calm         {{\cal M}}
\def\caln         {{\cal N}}
\def\calo         {{\cal O}}
\def\calp         {{\cal P}}

\def\reals        {{\mathbb R}}

\def\del          {\partial}

\def\tr           {\mathop{\rm Tr}}

\def\sqr#1#2{{\vcenter{\vbox{\hrule height.#2pt
 \hbox{\vrule width.#2pt height#1pt \kern#1pt
 \vrule width.#2pt}\hrule height.#2pt}}}}

\newcommand{\fft}[2]{{\frac{#1}{#2}}}
\newcommand{\ft}[2]{{\textstyle{\frac{#1}{#2}}}}
\def\jsquare{\mathop{\mathchoice{\sqr{8}{32}}{\sqr{8}{32}}
{\sqr{6.3}{9}}{\sqr{4.5}{9}}}}

\def\tQ{\tilde{Q}}
\def\a{\alpha}
\def\r{\rho}

\def\xh{\hat{x}}
\def\rh{\hat{\rho}}
\def\chih{\hat{\chi}}
\def\dd{\delta}
\def\ah{\hat{\alpha}}

\catcode`\@=12

\begin{document}


\title{Thermodynamics of the $\caln=2^*$ flow}

\pubnum{%
MCTP-03-25\\
hep-th/0305064}
\date{May 2003}

\author{
Alex Buchel and
James T. Liu\\[0.4cm]
\it Michigan Center for Theoretical Physics\\
\it Randall Laboratory of Physics, The University of Michigan\\
\it Ann Arbor, MI 48109-1120\\[0.2cm]
}

\Abstract{We discuss the thermodynamics of the $\caln=2^*$, $SU(N)$
gauge theory at large 't Hooft coupling. The tool we use is the
non-extremal deformation of the supergravity solution of Pilch and
Warner (PW) [hep-th/0004063], dual to $\caln=4$, $SU(N)$ gauge theory
softly broken to $\caln=2$.  We construct the exact non-extremal
solution in five-dimensional gauged supergravity and further uplift
it to ten dimensions.
Turning to the thermodynamics, we analytically compute the leading
correction in $m/T$ to the free energy of the non-extremal D3 branes
due to the PW mass deformation, and find that it is positive.  We
also demonstrate that the mass deformation of the non-extremal D3
brane geometry induces a temperature dependent gaugino condensate.
We find that the standard procedure of extracting the $\caln=2^*$
gauge theory thermodynamic quantities from the dual supergravity leads
to a violation of the first law of thermodynamics. We speculate on
a possible resolution of this paradox.}


\makepapertitle

\body

\version\versionno

\section{Introduction}

Over the last few years, gauge theory/string theory duality \cite{m9711}
(see \cite{a9905} for a review) has proven to be a very useful tool to
address nonperturbative questions in gauge theories.  Essentially,
this duality states that, say, four dimensional gauge theories at
large 't Hooft coupling, $g_{YM}^2 N\gg 1$, can be described by dual string
theories in weakly curved supergravity backgrounds. In the large
$N$ but fixed 't Hooft coupling limit, the string coupling vanishes,
and one can consistently restrict the string theory side of the
correspondence to the massless sector of type IIB supergravity.  Thus
the computations on the supergravity side shed light on the
nonperturbative gauge theory dynamics.  On the other hand,
nonperturbative effects in the gauge theory potentially could tell
us something new about the dual supergravity. This is indeed the
case with finite temperature phase transitions in gauge theories
\cite{w97,b0011,ghkt,bf,gtv}.  Specifically, in \cite{w97},
it was demonstrated how the (kinematic) confinement-deconfinement phase
transition of $\caln=4$, $SU(N)$ supersymmetric Yang-Mills theory
on $\reals\times S^3$ was mapped by the duality correspondence to the
previously discovered Hawking-Page phase transition in an anti-de Sitter
background \cite{hp}.

With a field theory intuition in mind, a new prediction for the
higher dimensional black holes (and a phase transition in an {\it
infinite volume}) was obtained in \cite{b0011}. Namely, using the
fact that the chirally symmetric phase of the Klebanov-Strassler
(KS) cascading gauge theories \cite{kt,ks} exists only above a
certain critical temperature, it was proposed in \cite{b0011} that
the dual supergravity backgrounds also would have a regular
Schwarzschild horizon only above certain horizon temperatures.  While
there is a good understanding of the high temperature thermodynamics
of the system (in a chirally symmetric phase) \cite{ghkt}, the low
temperature phase (and the phase transition) requires substantial
numerical work and is not yet understood.  Thus, strictly speaking,
the prediction for the new type of black holes of \cite{b0011} has
not yet been verified in the KS model.  Nevertheless, black holes
that exhibit a phase transition predicted in \cite{b0011} were shown
to exist \cite{bf,gtv} in a different (but closely related) model
--- the supergravity dual to pure $\caln=1$, $SU(N)$ SYM theory
\cite{mn}.  Unfortunately, the latter black holes were shown to be
thermodynamically unstable \cite{lst,gtv}%
\footnote{These black holes have a negative specific heat.},
and thus it is not clear whether the phase transition, while
mathematically allowed, is actually physically occurring.

In this paper we discuss yet another system which we argue undergoes
a finite temperature phase transition, namely the $\caln=2^*$ theory,
or in other words $\caln=4$, $SU(N)$ SYM softly broken by a
hypermultiplet mass term to $\caln=2$ gauge theory.
The supergravity dual to this gauge theory
was constructed by Pilch and Warner (PW) in \cite{pw}, and the
precise duality map between the gauge theory and the supergravity
was explained in \cite{bpp,cl}.  Here we consider the $\caln=2^*$
gauge theory at finite temperature, realized as a non extremal
deformation of the PW flow.  The argument for the existence of a
phase transition in this system is quite simple.  For the high
temperature phase, the mass deformation is irrelevant and we expect
the standard near-extremal D3 brane thermodynamics. In this phase
(which we refer to as the ``black hole phase'', $BH$) all the
thermodynamics quantities scale as $N^2$; for example, the entropy
goes as $S_{BH}\propto N^2 V T^3$.  On the other hand, the low
temperature phase is rather different. Imagine first completely
turning off the temperature.  In this case the $\caln=2^*$ gauge
theory is exactly soluble \cite{phil}. Its low energy effective
description (valid well below the hypermultiplet mass scale) is
given in terms of $\caln=2$, $U(1)^{N-1}$ gauge theory.  This gauge
theory has on the order of $N$ free degrees of freedom. We expect that
this low energy effective description is still valid for temperatures
much lower than the hypermultiplet mass. Thus in the low temperature
phase (which we denote the ``finite temperature Coulomb phase'' or
the Pilch-Warner phase, $PW$) the thermodynamic quantities of the
$\caln=2^*$ gauge theory would scale as the first power of $N$, so
that, {\it e.g.}, for the entropy, $S_{PW}\propto N V T^3$.  This
$N^2$ versus $N$ scaling of degrees of freedom suggests the presence
of a phase transition%
\footnote{We expect a phase transition in the strict $N\to\infty$ limit.
It is likely that at finite $N$ the phase transition is replaced by a
crossover regime.  We thank Arkady Vainshtein for a useful discussion
on this point.}
between the $BH$ and $PW$ phases, each of
which in principle exists at all temperatures.

The physical picture we have in mind for the phase transition is
as follows. Start with a high temperature $BH$ phase. In this phase
the free energy
will start negative at very high temperatures (as for the black D3
branes), and would gradually increase as the temperature is lowered.
We expect that at some $T=T_c$ the free energy would become zero,
$F_{BH}(T_c)=0$, and for $T<T_c$ it would become positive.  In the
large $N$ limit, the free energy of the $PW$ phase is down by ${\cal
O}(1/N)$, and may be taken to be zero. Thus the $PW$ phase should
be thermodynamically favorable for low temperatures, $T< T_c$. We
expect that both the $PW$ and the $BH$ phases are thermodynamically
stable. Moreover, there should be a well defined high temperature
expansion of the $BH$ phase%
\footnote{The high temperature expansion of the KS model developed
in \cite{ghkt} is ill defined in the ultraviolet.  This is related
to the unusual UV properties of the cascading gauge theories; for a
review see \cite{rks}.}.

In the next section we summarize our results.  In section 3 we
recall the salient features of the $\caln=2^*$ gauge theory, and
discuss the expectation for its finite temperature deformation. We
then review the PW $\caln=2^*$ renormalization group (RG) flow in
five dimensions and discuss its non-extremal deformation. The exact
ten dimensional lift of the ``temperature deformed'' five dimensional
PW flow is constructed in section 4.  In section 5 we analytically
determine the leading in $m/T$ correction to the near-extremal D3
brane geometry induced by the $\caln=2$ hypermultiplet mass $m$.
Finally, in section 6 we discuss the thermodynamics and the signature
of the phase transition. First, we explain the computation of the
free energy and the energy (mass) of the deformed supergravity
backgrounds, and compute the difference of the free energies of the
$BH$ and the $PW$ phases. This computation is valid for arbitrary
values of $m/T$, and thus can be used to study the phase transition.
Then, using the results of section 5, we compute the leading
correction (in the high temperature phase) to the black D3 brane
thermodynamics.
We find that the first law of thermodynamics applied to the
high temperature phase is violated. We speculate on
the relevance of the (induced) chemical potential for the
resolution of this paradox.
We end with some comments on the numerical
verification of the phase transition.

Before proceeding, we would like to comment on the study of the
thermodynamics of the closely related $\caln=1^*$, $SU(N)$ gauge
theory \cite{fm}.  As its name suggests, the $\caln=1^*$ gauge
theory is $\caln=4$, $SU(N)$ SYM softly broken by a chiral multiplet
mass term to $\caln=1$. Unlike the $\caln=2^*$ theory, however, the
dual supergravity background to this model \cite{ps} (PS) is known
only in the probe approximation. The study of thermodynamics in
\cite{fm} was also done in the probe approximation.  Only the entropy
was computed in the high temperature regime of the non-extremal PS
background;
while the free energy and the energy were not computed independently,
they were obtained by enforcing the first law of thermodynamics.
As we will see from the thermodynamics of the $\caln=2^*$
model discussed here, the computation of the entropy alone does not
allow us to reproduce the free energy --- it appears
one needs to compute the
induced chemical potential as well.  It would be of interest to
repeat the $\caln=2^*$ analysis presented here to the $\caln=1^*$
model.  But first, the exact extremal geometry of the $\caln=1^*$
theory has to be understood.

\section{Summary of results and outlook}

As the bulk of the paper is rather technical, we highlight our main
results in this section.
\nxt
We first observe that the five dimensional gauged supergravity flow
of $\caln=2^*$ PW \cite{pw} can be deformed to yield a non-extremal
black hole geometry with regular horizon.  The consistency of the
$D=5$, $\caln=8$ gauged supergravity truncation then implies that
this black hole solution can be uplifted to the full ten dimensional
solution of type IIB supergravity. We explicitly verify that this
is indeed so.  This non-extremal deformation is interpreted as the
supergravity dual to the finite temperature $\caln=2^*$ $SU(N)$
gauge theory in the deconfined phase, which we refer to as the $BH$
phase.
\nxt
We show that there is a three parameter family of five dimensional
black holes admitting regular horizons.  These three parameters are
the temperature $T$ and the (generically different) masses of the
bosonic $m_b$ and fermionic $m_f$ components of the $\caln=2$
hypermultiplet.  All regular horizon non-extremal solutions asymptote
to $AdS_5$, which is consistent with the gauge theory expectation
that both the temperature and the mass deformations should be
irrelevant in the ultraviolet of the gauge theory. Asymptotic
$\caln=2$ supersymmetry of the extremal PW geometry imposes a
constraint on the leading nontrivial asymptotics of the two five
dimensional  supergravity scalars in the non-extremal deformation.
The latter reduces the number of independent parameters of the
regular horizon solution to two: one related to the temperature,
and the other to the $\caln=2$ hypermultiplet mass $m=m_b=m_f$.
\nxt
In the high temperature limit, $\a_1\equiv (m_b/T)^2\ll 1$, $\a_2\equiv
m_f/T\ll 1$, the five dimensional black hole solution is a small
deformation of the finite temperature $AdS_5$ geometry, representing
the $S^5$ reduction of the throat region of the near extremal D3
branes. We analytically determine the leading correction in $\a_i$
of the near extremal $AdS_5$ geometry. As expected from gauge theory
arguments, asymptotic $\caln=2$ supersymmetry sets  $\a_1\sim
(\a_2)^2$.
\nxt
After constructing the black hole solution, we turn to the study
of thermodynamics.  We discuss the computation of the free energy
$F$, the entropy $S$, and the energy $E$ of the non-extremal deformation
of the PW flow.  The entropy is just the Bekenstein-Hawking entropy
of the horizon, and is determined from the infrared data of the
geometry.  The free energy, or more precisely $F T$, is the Euclidean
gravitational action, and the energy $E$ is the conserved ADM mass
of the geometry.  Note that computing both $E$ and $F$ requires the
knowledge of the IR and the UV data of the solution.  Furthermore,
we verify that $F=E- T S$ is identically satisfied in the supergravity.
\nxt
While the computation of the entropy is straightforward, both the
free energy and the energy diverges, and requires regularization.
Following \cite{HH}, we compute $F$ and $E$ with respect to a
reference geometry which we take to be the supersymmetric PW flow
with periodically identified (Euclidean) time direction with
periodicity equal to the inverse horizon temperature. We call this
geometry the ``$PW$ phase''.  The prescription of \cite{HH} requires
the introduction of a boundary cutoff and the matching of induced
geometries (and matter fields) for the background at hand and the
reference one ``up to sufficiently high order'' \cite{HH}.  We apply
the ``minimal subtraction'' prescription for matching, where only
the leading asymptotics of the induced geometries and the matter
fields are matched. This prescription gives the correct answers for
simple black hole geometries such as the Schwarzschild-anti-de
Sitter solution. It also works in more complicated cases such as
the nonabelian black hole solutions of \cite{gtv}.
\nxt
Using this minimal subtraction prescription for the free energy and
the energy, we find an explicit analytical expression for $F_{BH}-F_{PW}$
(or $E_{BH}-E_{PW}$) is terms of the coefficients of the subleading
ultraviolet asymptotics of the five dimensional scalars inducing
the Pilch-Warner flow \cite{pw}.  An added bonus of using the PW
background
(with appropriately compactified Euclidean time direction)
as the reference one in $F$ and $E$ regularization is the fact that
the purported phase transition between the high temperature $BH$
phase and the low temperature $PW$ phase arises when
$\Delta^{BH}_{PW}\equiv F_{BH}-F_{PW}$ changes sign.
\nxt
Using the high temperature expansion (corresponding to deformations from
the non-extremal $AdS_5$ geometry), and working to leading order in $m/T$,
we analytically compute the corresponding deformations of the thermodynamic
quantities.  While $F=E-TS$ continues to be satisfied after regulation, we
however find that $T dS \ne dE$.
We have verified our prediction for the leading correction to the free
energy numerically.  This indirectly confirms the violation of the first
law of thermodynamics.
\nxt
Though we have been unable to find a satisfactory explanation
for this apparent contradiction with the first law of thermodynamics
for the high temperature phase of the $\caln=2^*$ flow,
we point out that this paradox could be resolved
once we include a certain chemical potential induced by the fermionic
mass term of the $\caln=2$ hypermultiplet.

Perhaps the most intriguing conclusion we have reached is the fact
that the proper interpretation of a finite temperature deformation
of the Pilch-Warner geometry appears to require  the introduction of a
nonvanishing chemical potential dual to the sources that are turned
on in the UV.  This induced chemical potential follows
from the conjecture that the string theory partition function in
the gauge/string theory correspondence is dual to the gauge theory
grand canonical partition function. In section 6.3 we outline the
general arguments leading to such a statement.  It would be interesting
to verify this in a more general setting, {\it e.g.}, by studying the
finite temperature deformations of generic holographic renormalization
group flows as in \cite{sk1,sk2}.

The original motivation for the study of the $\caln=2^*$ thermodynamics
presented here was to study and confirm the phase transition
between the $BH$ and the $PW$ phases.  However, our analysis
for this question is as yet inconclusive.  While the high temperature
expansion is suggestive that such a phase transition occurs, additional
analytical or numerical work is required to extrapolate to the region
$m\sim T$ where solid evidence of the transition would be obtained.
We hope to report on these results in a separate publication.
Finally, it is interesting to understand the ``hydrodynamic
description'' of the $BH$ phase of the $\caln=2^*$ gauge theory
along the lines of \cite{st}.

\section{$\caln=2^*$ RG flow and its non-extremal deformation in
five dimensions}

\subsection{The gauge theory picture}

In the language of four-dimensional $\caln=1$ supersymmetry, the
mass deformed $\caln=4$ $SU(N)$ Yang-Mills theory ($\caln=2^*$) in
$\reals^{3,1}$ consists of a vector multiplet $V$, an adjoint chiral
superfield $\Phi$ related by $\caln=2$ supersymmetry to the gauge
field, and two additional adjoint chiral multiplets $Q$ and $\tilde{Q}$
which form an $\caln=2$ hypermultiplet.  In addition to the usual
gauge-invariant kinetic terms for these fields%
\footnote{The classical K\"{a}hler potential is normalized
according to $(2/g_{YM}^2)\tr[\bar{\Phi}\Phi+ \bar{Q}Q+\bar{\tQ}\tQ]$.},
the theory has additional interactions and a hypermultiplet mass term
given by the superpotential
\begin{equation}
W=\frac{2\sqrt{2}}{g_{YM}^2}\tr([Q,\tQ]\Phi)
+\frac{m} {g_{YM}^2}(\tr Q^2+\tr\tQ^2)\,.
\eqlabel{sp}
\end{equation}
When $m=0$ the gauge theory is superconformal with $g_{YM}$
characterizing an exactly marginal deformation. The theory has a
classical $3(N-1)$ complex dimensional moduli space, which is
protected by supersymmetry against (non)-perturbative quantum
corrections.

When $m\ne 0$, the $\caln=4$ supersymmetry is softly broken to
$\caln=2$. This mass deformation lifts the $\{Q,\ \tQ\}$ hypermultiplet
moduli directions, leaving the $(N-1)$ complex dimensional Coulomb
branch of the $\caln=2$, $SU(N)$ Yang-Mills theory, parameterized by
expectation values of the adjoint scalar
\begin{equation}
\Phi={\rm diag} (a_1,a_2,\cdots,a_N)\,,\quad \sum_i a_i=0\,,
\eqlabel{adsc}
\end{equation}
in the Cartan subalgebra of the gauge group.  For generic values
of the moduli $a_i$, the gauge symmetry is broken to that of the
Cartan subalgebra $U(1)^{N-1}$, up to the permutation of individual
$U(1)$ factors. Additionally, the superpotential \eqref{sp} induces
the RG flow of the gauge coupling.  While from the gauge theory
perspective it is straightforward to study this $\caln=2^{*}$ theory
at any point on the Coulomb branch \cite{phil}, the PW supergravity
flow \cite{pw} corresponds to a particular Coulomb branch vacuum.
More specifically, matching the probe computation in gauge theory
and the dual PW supergravity flow, it was argued in \cite{bpp} that
the appropriate Coulomb branch vacuum corresponds to a  linear
distribution of the vevs \eqref{adsc} as
\begin{equation}
a_i\in [-a_0,a_0],\qquad a_0^2=\frac{m^2 g_{YM}^2 N}{\pi}\,,
\eqlabel{inter}
\end{equation}
with (continuous in the large $N$ limit) linear number density
\begin{equation}
\rho(a)=\frac{2}{m^2 g_{YM}^2}\sqrt{a_0^2-a^2},\qquad
\int_{-a_0}^{a_0}da\,\rho(a)=N\,.
\eqlabel{rho}
\end{equation}
Unfortunately, the extension of the $N=2^*$ gauge/gravity
correspondence of \cite{pw,bpp,cl} for vacua other than \eqref{rho}
is not known.

In \cite{bpp,cl} the dynamics of the gauge theory on the D3 brane
probe in the PW background was studied in detail.  It was shown
in \cite{bpp} that the probe has a one complex dimensional moduli
space, with bulk induced  metric precisely equal to the metric on
the appropriate one complex dimensional submanifold of the $\caln=2^*$,
$SU(N+1)$ Donagi-Witten theory Coulomb branch.  This one dimensional
submanifold is parameterized by the expectation value $u$ of the
$U(1)$ complex scalar on the Coulomb branch of the theory where
$SU(N+1)\rightarrow U(1)\times SU(N)_{PW}$. Here the $_{PW}$ subscript
denotes that the $SU(N)$ factor is in the Pilch-Warner vacuum
\eqref{rho}. Whenever $u$ coincides with any of the $a_i$ of the
PW vacuum, the moduli space metric diverges, signaling the appearance
of additional massless states.  An identical divergence is observed
\cite{bpp,cl} for the probe D3-brane at the {\it enhan\c{c}on}
singularity of the PW background.  Away from the singularity locus,
$u=a\in [-a_0,a_0]$, the gauge theory computation of the probe
moduli space metric is 1-loop exact.  This is due to the suppression
of instanton corrections in the large $N$ limit \cite{bpp,b} of
$\caln=2$ gauge theories.

Consider now $\caln=2^*$ gauge theory at finite temperature $T$.
Turning on a mass $m$ for the hypermultiplet sets a strong coupling
scale $\Lambda \propto m$.  We expect to find two different phases
of this gauge theory, depending on whether $T\gg \Lambda$ or $T\ll
\Lambda$.  In the former case the effect of the mass deformation
is negligible, and we expect to recover the $\caln=4$ thermodynamics.
In particular, conformal invariance dictates that the free energy
scales like $T^4$, with a prefactor of $N^2$ indicative of the
scaling of the number of degrees of freedom.  At weak 't Hooft
coupling the familiar result reads \cite{gkp}
\begin{equation}
F_{SYM}=-\frac{\pi^2}{6} N^2 V T^4.
\eqlabel{fsym}
\end{equation}
By symmetry arguments, we expect the corrections to the free energy
\eqref{fsym} due to the mass deformation \eqref{sp} to be of order
${\cal O}\left(m^2/T^2\right)$.  The $N^2$ scaling of the thermodynamic
quantities naturally occurs in the ten-dimensional black holes
describing the non-extremal deformation of the dual supergravity
backgrounds. For this reason we will call the high temperature phase
of $\caln=2^*$ SYM the $BH$ or black hole phase.

\begin{figure}[t]
\begin{center}
\epsfig{file=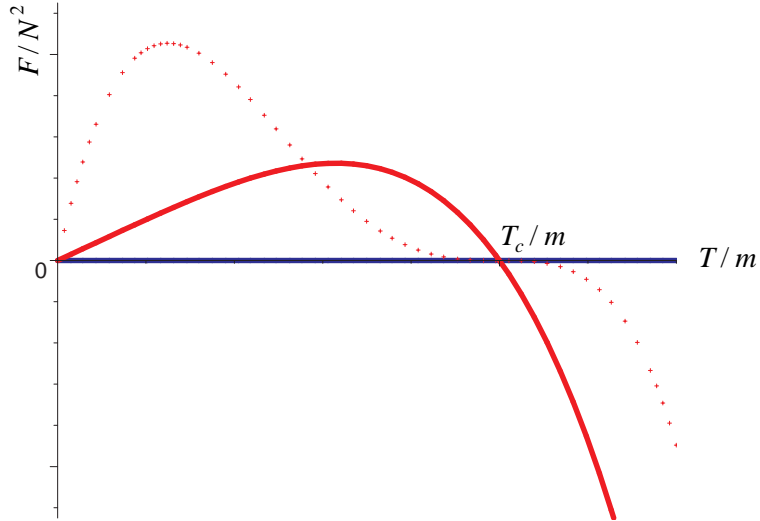,width=0.7\textwidth}
\caption{The expected behavior of free energies of the $BH$ (red curves)
and the $PW$ (blue curve) phases ($F_{BH}$ and $F_{PW}$) as a function of
$T/m$ in the large $N$ limit.  The dotted line for the $BH$ phase
would imply a second order phase transition, while the solid line would
imply a first order phase transition.}
\label{cases}
\end{center}
\end{figure}

In the other limit, we expect qualitatively different physics in
the low temperature phase of $\caln=2^*$ SYM. Ignoring $T$ in the
first approximation, the low energy effective description of the
$\caln=2^*$ theory is given by free $U(1)^{N-1}$ SYM as explained
above.  This effective description breaks down at scales of order
the strong coupling scale, \ie\ $m$, but is appropriate as we turn
on the temperature provided $T\ll m$.  The number of (free) degrees
of freedom of this effective low energy description scales like
$N$, which must be reflected in the scaling of the thermodynamics
quantities such as the entropy, $S\propto N V T^3 $. We denote this
phase the ``finite temperature Coulomb phase'', or the $PW$ phase.
Notice that
\begin{equation}
\frac{F_{PW}}{F_{BH}}\sim \frac 1N,
\eqlabel{fsymr}
\end{equation}
and thus vanishes in the large-$N$ limit.  Qualitatively, we expect the
free energy of the $\caln=2^*$ SYM to behave as in Fig.~\ref{cases}.

Given two possible phases, $BH$ and $PW$, the most favorable one
is that with lowest free energy.  Thus the signature of a phase
transition in the $\caln=2^*$ theory would be a change in sign in
the difference $\Delta^{BH}_{PW}\equiv F_{BH}-F_{PW}$.  This
suggests that, in order to have a phase transition, the sign of the
$O(m^2/T^2)$ correction to \eqref{fsym} must be positive so that,
when the temperature is lowered from an initial high temperature
phase, $F_{BH}$ will be driven above $F_{PW}$ at a finite temperature
$T_c$.  While we have not performed this computation perturbatively
in the 't Hooft coupling, we have instead analytically determined
the ${\cal O}(m^2/T^2)$ correction for strong 't Hooft coupling in
the dual supergravity computation. We find that this correction
indeed has a positive coefficient \eqref{thf0} at strong 't Hooft
coupling.  This provides evidence for a phase transition, although
further investigation is necessary beyond leading order to substantiate
this claim.

\subsection{The PW renormalization group flow}

The gauge theory RG flow induced by the superpotential \eqref{sp}
corresponds to a five dimensional gauged supergravity flow induced
by a pair of scalars,  $\a\equiv \sqrt{3}\ln \r$ and $\chi$.
The effective five-dimensional action is
\begin{equation}
S=\frac{1}{4\pi G_5}\,
\int_{\calm_5} d\xi^5 \sqrt{-g}\left(\ft14 R-(\del\a)^2-(\del\chi)^2-
\calp\right)\,,
\eqlabel{action5}
\end{equation}
where the potential $\calp$ is%
\footnote{We set the 5d gauged supergravity coupling to one. This
corresponds to setting the $S^5$ radius $L=2$.}
\begin{equation}
\calp=\frac{1}{16}\left[ \left(\frac{\del W}{\del \a}\right)^2+
\left(\frac{\del W}{\del \chi}\right)^2\right]-\frac 13 W^2\,,
\eqlabel{pp}
\end{equation}
with the superpotential
\begin{equation}
W=-\frac{1}{\r^2}-\frac 12 \r^4 \cosh(2\chi)\,.
\eqlabel{supp}
\end{equation}
Note that we have chosen identical normalizations for $\alpha$ and $\chi$
to highlight the general $D=5$, $\caln=2$ features of the flow.  The PW
geometry \cite{pw} has the flow metric
\begin{equation}
ds_5^2=e^{2 A} \left(-dt^2 +d\vec{x}\,^2\right)+dr^2\,.
\eqlabel{flowmetric}
\end{equation}
Solving the Killing spinor equations for a supersymmetric flow then yields
the first order equations
\begin{equation}
\frac {d A}{d r}=-\frac 13 W\,,\qquad
\frac {d\a}{d r}=\frac{1}{4} \frac{\del W}{\del \a}\,,\qquad
\frac {d\chi}{d r}=\frac{1}{4} \frac{\del W}{\del \chi}\,.
\eqlabel{pwo}
\end{equation}
It is straightforward to verify that solutions to these flow equations
will automatically satisfy the scalar and Einstein equations of motion.

\subsubsection{Asymptotics of the PW flow}

This system was solved in \cite{pw} by rewriting the equations in terms of
$\chi$ as an independent variable.  Given this explicit solution of the
flow equations, it is easy to extract the UV/IR asymptotics.
In the ultraviolet, $r\to +\infty$, we find
\begin{equation}
{\rm UV:}\qquad\qquad \rho\to 1_-,\qquad \chi\to 0_+,\qquad A\to \frac 12
\ r\,.
\eqlabel{pwasi}
\end{equation}
This corresponds to the scalars approaching the maximally symmetric
$AdS_5\times S^5$ UV fixed point of the potential ${\cal P}$.  In the
infrared, $r\to 0$, we find instead
\begin{equation}
{\rm IR:}\qquad\qquad \rho\to 0_+,\qquad \chi\to +\infty,\qquad
A\to -\frac 83 \chi\,.
\eqlabel{pwas0}
\end{equation}
As will be apparent later, this flow to the IR will be cut off at finite
temperature.

\subsection{The non-extremal PW flow}

We now consider deforming the PW flow by turning on non-extremality in the
metric.  Since the deformed flow breaks supersymmetry, we can no longer
appeal to first order equations, but must consider the second order
equations of motion. The action \eqref{action5} yields the Einstein equation
\begin{equation}
\fft14 R_{\mu\nu}=\del_\mu \a \del_\nu\a+\del_\mu\chi\del_\nu\chi
+\fft13 g_{\mu\nu} \calp\,,
\eqlabel{ee}
\end{equation}
and the scalar equations
\begin{equation}
\jsquare\alpha=\fft12\fft{\del\calp}{\del\alpha}\,,\qquad
\jsquare\chi=\fft12\fft{\del\calp}{\del\chi}\,.
\eqlabel{scalar}
\end{equation}
For a finite temperature deformation of the flow metric \eqref{flowmetric},
we take
\begin{equation}
ds_5^2=e^{2 A} \left(-e^{2 B}\ dt^2 +d\vec{x}\,^2\right)+dr^2\,,\\
\eqlabel{ab}
\end{equation}
where $e^{2B}$ represents a blackening function.  Note that we choose to
retain $g_{rr}=1$ since any non-trivial factor can be absorbed into a
redefinition of $r$.

Substituting this metric ansatz into the equations of motion,
\eqref{ee} and \eqref{scalar}, we find
\begin{equation}
\begin{split}
0&=\a''+\left(4 A' + B'\right)\a' -\frac 12 \frac{\del\calp}{\del\a}\,,\\
0&=\chi''+\left(4 A' + B'\right)\chi' -\frac 12 \frac{\del\calp}{\del\chi}
\,,\\
0&=B''+\left(4 A'+B'\right) B'\,,\\
&\frac 14 A''+\frac 14 B''+
\left(A'\right)^2+\frac 14\left(B'\right)^2+\frac 54 A' B'=-\frac 13 \calp\,,\\
&- A''-\frac 14  B''-\left(A'\right)^2
-\frac 14 \left(B'\right)^2-\frac 12 A' B'
=\left(\a'\right)^2+\left(\chi'\right)^2 +\frac 13\calp\,.
\end{split}
\eqlabel{aeq}
\end{equation}
Note that, by defining $\bar A\equiv A+\fft14B$ and $\bar B\equiv \sqrt{3}B/4$
and taking appropriate linear combinations, the above equations may be
written in the equivalent form
\begin{equation}
\begin{split}
0&=\a''+4\bar A'\a'-\fft12\fft{\del\calp}{\del\a}\,,\\
0&=\chi''+4\bar A'\chi'-\fft12\fft{\del\calp}{\del\chi}\,,\\
0&=\bar B''+4 \bar A' \bar B'-\fft12\fft{\del\calp}{\del\bar B}\,,\\
3(\bar A')^2&=(\a')^2+(\chi')^2+(\bar B')^2-\calp\,,\\
3\bar A''&=-4\left((\a')^2+(\chi')^2+(\bar B')^2\right)\,,
\end{split}
\eqlabel{neweom}
\end{equation}
where we have formally introduced $\del\calp/\del\bar B\equiv0$.  When
written in this form, it is easy to see that the last equation is
redundant, and may be obtained by differentiating the penultimate equation
and substituting back in the scalar equations of motion.  Since these
equations are consistent, we can use the same scalars as in the PW case,
even when considering deformed flows.

At this point, a few comments are in order.  Firstly, the last equation of
\eqref{neweom} provides a non-extremal generalization of the holographic
$c$-theorem, namely $\bar A''\le0$ where $e^{4\bar A}=\sqrt{-g}$.
Secondly, scalars in $AdS_5$ may be labeled by the representations
$D(E_0,0,0)$ where $E_0$ is the lowest energy state, and may be related
to the conformal dimension, $\Delta$, of the dual field theory operators.
Expansion of $\calp$ about the UV fixed point indicates that $\a$ (or
$\rho$) has $E_0=2$, while $\chi$ has $E_0=3$.  The blackening factor
$\bar B$ may be thought of as a scalar mode with $E_0=4$.  Finally, we see
that the equation for $B$ in \eqref{aeq} can be integrated once to obtain
\begin{equation}
\ln B' +4 A +B\ =\ {\rm const},
\eqlabel{intB}
\end{equation}
or equivalently $\ln B'+4\bar A=\ {\rm const}$.  This relation will prove
useful below.

\subsubsection{Asymptotics of the finite temperature deformation}

The thermal solutions we are interested in have regular black hole horizons.
As a result, we may examine the solution to the system of equations
\eqref{aeq} near the horizon.  In this case, the behavior of \eqref{pwas0}
is cut off, and the scalars run to fixed values, $\rho_0$ and $\chi_0$,
on the horizon.  Including the horizon value of $A$, we see that the
nonsingular in the IR flows are given by a three parameter family
$\{\alpha,\r_0>0,\chi_0\}$, specifying the near horizon ($r\to0$)
Taylor series expansions
\begin{equation}
\begin{split}
e^A&=e^{\a}\,
\left[1+\left(\sum_{i=1}^{\infty}\ a_i\ r^{2 i}\right)\right]\,,\\
e^B&=\delta\ r\left(1+\sum_{i=1}^{\infty}\ b_i\ r^{2 i} \right)\,,\\
\r&=\r_0+\left(\sum_{i=1}^{\infty}\ \r_i\ r^{2 i}\right)\,,\\
\chi&=\chi_0+\left(\sum_{i=1}^{\infty}\ \chi_i\ r^{2 i}\right)\,.
\end{split}
\eqlabel{ds0asa}
\end{equation}
Here, $\delta=\delta(\r_0,\chi_0)$ should be adjusted so that
$e^B\to 1_-$ as $r\to +\infty$.  The first non-trivial terms in the
series expansions \eqref{ds0asa} are
\begin{equation}
\begin{split}
\dd^{-2}\ a_1&=\ft{1}{12}\ \r_0^{-4}+\ft{1}{6}\ \r_0^2\ \cosh(2\chi_0)-
\ft{1}{48}\ \r_0^8\ \sinh^2(2\chi_0)\,,\\
\dd^{-2}\ b_1&=-\ft{1}{9}\ \r_0^{-4}-\ft{2}{9}\ \r_0^2\ \cosh(2\chi_0)+
\ft{1}{36}\ \r_0^8\ \sinh^2(2\chi_0)\,,\\
\dd^{-2}\ \r_1&=\ft{1}{24}\ \r_0^{-3}-\ft{1}{24}\ \r_0^3\ \cosh(2\chi_0)+
\ft{1}{48}\ \r_0^9\ \sinh^2(2\chi_0)\,,\\
\dd^{-2}\ \chi_1&=-\ft{1}{8}\ \r_0^{2} \sinh(2\chi_0)
    +\ft{1}{64}\ \r_0^8\ \sinh(4\chi_0)\,.
\end{split}
\eqlabel{1dsa}
\end{equation}
%


\section{The ten-dimensional solutions}

In this section, we lift the deformed PW flow to ten dimensions.  However
before doing so we establish our conventions and review some of the
pertinent aspects of the lifting procedure.

\subsection{Type IIB supergravity equations of motion}

We use a mostly positive convention for the signature $(-+\cdots +)$
and take $\epsilon_{1\cdots10}=+1$.  The type bosonic IIB equations
consist of the following \cite{schwarz83}:

\noindent $\bullet$\quad  The Einstein equations:
\begin{equation}
R_{MN}=T^{(1)}_{MN}+T^{(3)}_{MN}+T^{(5)}_{MN}\,,
\eqlabel{tenein}
\end{equation}
where the energy momentum tensors of the dilaton/axion field, $\calb$,
the three index antisymmetric tensor field, $F_{(3)}$, and the self-dual
five-index tensor field, $F_{(5)}$, are given by
\begin{equation}
T^{(1)}_{MN}= P_MP_N{}^*+P_NP_M{}^*\,,
\eqlabel{enmomP}
\end{equation}
\begin{equation}
T^{(3)}_{MN}=
       \frac 18(G^{PQ}{}_MG^*_{PQN}+G^{*PQ}{}_MG_{PQN}-
        \frac 16 g_{MN} G^{PQR}G^*_{PQR})\,,
\eqlabel{enmomG}
\end{equation}
and
\begin{equation}
T^{(5)}_{MN}= \frac 16 F^{PQRS}{}_MF_{PQRSN}\,.
\eqlabel{enmomF}
\end{equation}
In the unitary gauge, $\calb$ is a complex scalar field, and
\begin{equation}
P_M= f^2\partial_M \calb\,,\qquad Q_M= f^2\,{\rm Im}\,(
\calb\partial_M\calb^*)\,,
\eqlabel{defofPQ}
\end{equation}
where
\begin{equation}
f= \frac{1}{ (1-\calb \calb^*)^{1/2}}\,,
\eqlabel{defoff}
\end{equation}
while the antisymmetric tensor field $G_{(3)}$ is given by
\begin{equation}
G_{(3)}= f(F_{(3)}-\calb F_{(3)}^*)\,.
\eqlabel{defofG}
\end{equation}

\noindent
$\bullet$\quad The Maxwell equations:
\begin{equation}
(\nabla^P-i Q^P) G_{MNP}= P^P G^*_{MNP}-\frac 23\,i\,F_{MNPQR}
G^{PQR}\,.
\eqlabel{tenmaxwell}
\end{equation}

\noindent
$\bullet$\quad The dilaton equation:
\begin{equation}
(\nabla^M -2 i Q^M) P_M= -\frac{1}{ 24} G^{PQR}G_{PQR}\,.
\eqlabel{tengsq}
\end{equation}

\noindent
$\bullet$\quad The self-dual equation:
\begin{equation}
F_{(5)}= \star F_{(5)}\,.
\eqlabel{tenself}
\end{equation}

\noindent
In addition, $F_{(3)}$ and $F_{(5)}$ satisfy Bianchi identities which
follow from the definition of the field strengths in terms of their
potentials:
\begin{equation}
\begin{split}
F_{(3)}&= dA_{(2)}\,,\\
F_{(5)}&= dA_{(4)}-{\frac 18}\,{\rm Im}( A_{(2)}\wedge
F_{(3)}^*)\,.
\end{split}
\eqlabel{defpotth}
\end{equation}

For the ten-dimensional uplift of the RG flows in the five-dimensional
gauged supergravity, the metric ansatz and the dilaton is basically
determined by group theoretical properties of the $D=5$, $\caln=8$
scalars.  Thus they must be the same for both the deformed and original
PW flows.  Specifically, we assume \cite{pw} that the $D=10$ Einstein
frame metric is
\begin{equation}
\begin{split}
ds_{10}^2&=\Omega^2 ds_5^2 + 4 \frac {(c X_1 X_2)^{1/4} }{\r^3}\biggl(
c^{-1} d\theta^2+\r^6\cos^2\theta \left(\frac {\sigma_1^2}{c X_2}
+\frac{\sigma_2^2+\sigma_3^2}{X_1}\right)+\sin^2\theta\frac {d\phi^2}{X_2}
\biggr)\,,
\end{split}
\eqlabel{10m}
\end{equation}
where $ds_5^2$ is either the original PW flow metric
\eqref{flowmetric} or its deformations \eqref{ab},
and $c\equiv \cosh (2\chi)$. The warp factor is given by
\begin{equation}
\Omega^2=\frac {(c X_1 X_2)^{1/4} }{\r}\,,
\eqlabel{om5}
\end{equation}
and the two functions $X_i$ are defined by
\begin{equation}
\begin{split}
X_1(r,\theta)&=\cos^2\theta+\r(r)^6\cosh(2\chi(r))\sin^2\theta\,,\\
X_2(r,\theta)&=\cosh(2\chi(r))\cos^2\theta+\r(r)^6\sin^2\theta\,.
\end{split}
\eqlabel{x1x2}
\end{equation}
As usual, $\sigma_i$ are the $SU(2)$ left-invariant forms normalized so that
$d\sigma_i=2 \sigma_j\wedge \sigma_k$.
Note that we perform all computations in the natural orthonormal frame
given by
\begin{equation}
\begin{split}
&e^1\propto dt,\quad e^2\propto dr,\quad e^3\propto dx_1,
\quad e^4\propto dx_2,\quad e^5\propto dx_3,\\
&e^6\propto d\theta,\quad e^7\propto \sigma_1,
\quad e^8\propto \sigma_2,\quad e^9\propto \sigma_3,\quad
e^{10}\propto  d\phi\,,
\end{split}
\eqlabel{frame}
\end{equation}

Turning now to the matter fields, for the dilaton/axion we have
\begin{equation}
f=\frac 12 \left(\left(\frac{c X_1 }{X_2}\right)^{1/4}+
\left(\frac{c X_1 }{X_2}\right)^{-1/4}\right),\qquad
f\calb =\frac 12 \left(\left(\frac{c X_1 }{X_2}\right)^{1/4}-
\left(\frac{c X_1 }{X_2}\right)^{-1/4}\right) e^{2i \phi}\,.
\eqlabel{dilax}
\end{equation}
The consistent truncation ansatz does not specify the 3-form nor 5-form
fluxes. As in \cite{pw}, for the 2-form potential we assume the most
general ansatz allowed by the global symmetries of the background
\begin{equation}
A_{(2)}=e^{i\phi}\bigl(a_1(r,\theta)\ d\theta\wedge \sigma_1+a_2(r,\theta)\
 \sigma_2
\wedge \sigma_3+a_3(r,\theta)\ \sigma_1\wedge d\phi+a_4(r,\theta)\
d\theta\wedge d\phi\bigr)\,,
\eqlabel{a2}
\end{equation}
where $a_i(r,\theta)$ are arbitrary complex functions.
For the 5-form flux we assume
\begin{equation}
F_5=\calf+\star\calf,\qquad
\calf=dt\wedge {\rm vol}_{\reals^3}\wedge d\omega\,,
\eqlabel{5form}
\end{equation}
where $\omega(r,\theta)$ is an arbitrary function.
As in the PW case, examination of the Einstein equations
reveals that 2-form potential functions $a_i$ have the following
properties: $a_4\equiv 0$; $a_1$, $a_2$ are pure imaginary, and $a_3$
is real.

\subsection{Lift of the near extremal deformation}

The verification of the uplifted solution proceeds
exactly as for the $\caln=2^*$ flow deformations
discussed in \cite{n2def}.
Thus we present only the results.   We find
\begin{equation}
\begin{split}
a_1&=- i\ 4\ \tanh(2\chi) \cos\theta\,, \\
a_2&=i\ 4\ \frac{\r^6\sinh(2\chi)}{X_1}\ \sin\theta\cos^2\theta\,,\\
a_3&=-4\  \frac{\sinh(2\chi)}{X_2}\ \sin\theta\cos^2\theta\,,
\end{split}
\eqlabel{aaa}
\end{equation}
and
\begin{equation}
\begin{split}
&\frac{\del \omega}{\del \theta}=-\frac 32 e^{4 A + B} \left(\ln \r\right)'\
\sin 2\theta\,,\\
&\frac{\del \omega}{\del r}=\frac 18 e^{4 A + B}\
\frac{1}{\r^4}\ \biggl(-\r^{12}\sinh^2(2\chi)\sin^2\theta +2\r^6
\cosh(2\chi)(1+\sin^2\theta)+2\cos^2\theta\biggr)\,.
\end{split}
\eqlabel{wres}
\end{equation}
We have explicitly verified that by supplementing the metric and the
dilaton/axion ansatz of the previous section with \eqref{aaa},
\eqref{wres} and the five-dimensional flow equations \eqref{aeq},
all the equations of ten-dimensional type IIB supergravity are satisfied.


\section{High temperature expansion}

Having examined the system of equations governing the non-extremal flow,
\eqref{aeq}, we now turn to the construction of solutions.  At finite
temperatures, we find it convenient to parametrize the flow not in terms
of the radial coordinate $r$, but rather in terms of the blackening
function $e^B$.  To do so, we introduce a new coordinate
\begin{equation}
y\equiv e^{B},\qquad y\in[0,1]\,,
\eqlabel{y}
\end{equation}
with $y=0$ being the horizon and $y\to 1_-$ the UV asymptotic limit.
The standard near-extremal D3 brane solution is realized when the bosonic
and fermionic masses of the $\caln=2$ hypermultiplet components are
turned off, corresponding to the supergravity scalars $\rho$ and $\chi$
sitting at the UV fixed point.  The near-extremal D3 brane solution has
the form
\begin{equation}
\begin{split}
A(y)&=\ah-\frac 14 \ln (1-y^2)\,,\\
\r(y)&=1\,,\\
\chi(y)&=0\,,
\end{split}
\eqlabel{d3non}
\end{equation}
where $\ah$ is an integration constant which determines the
BH temperature according to
\begin{equation}
T=\frac {1}{2\pi}\, e^{\ah}\,.
\eqlabel{tem}
\end{equation}
%

\begin{figure}
\begin{center}
\epsfig{file=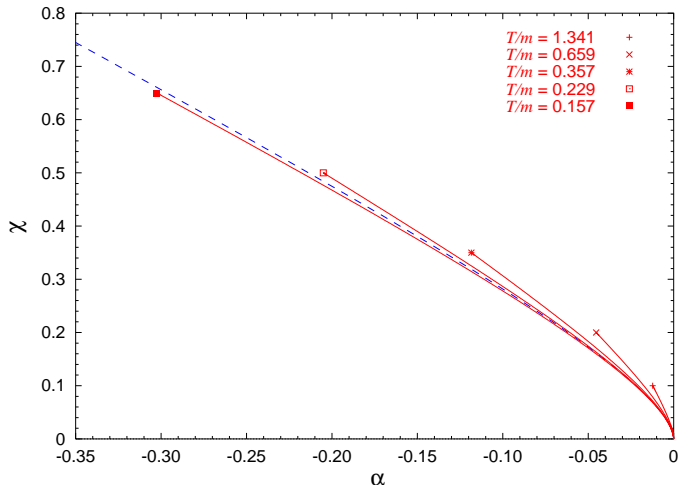,width=0.6\textwidth}
\caption{Finite temperature PW flows away from the UV stable fixed point
$(\alpha,\chi)=(0,0)$.  The endpoints of the flows correspond to
the fixed values of the scalars at the horizon.  The dashed line gives the
supersymmetric zero-temperature flow.}
\label{fig:flow}
\end{center}
\end{figure}

We recall that at zero temperature the PW flow involves the scalars
$\rho$ and $\chi$ running away from the UV fixed point as one flows to
the IR.  At a regular horizon, the scalars attain fixed values $\rho_0$
and $\chi_0$.  Hence we now seek a solution to \eqref{aeq} satisfying
the conditions
\begin{equation}
(\rho,\chi)\to(1,0)\quad\hbox{as}\quad y\to1_-\,,\qquad
(\rho,\chi)\to(\rho_0,\chi_0)\quad\hbox{as}\quad y\to0\,.
\end{equation}
Several flows satisfying these boundary conditions are displayed in
Fig.~\ref{fig:flow}.  It is clear from the figure that the flows proceed
further into the IR as the temperature is lowered.
While we have been unable to find an exact analytical solution, it is
nevertheless possible to develop a consistent (uniformly convergent)
perturbative approximation in the high temperature phase. On the
gauge theory side, this corresponds to a power series expansion in
$\a_1\propto (m_b)^2\ll 1$ and $\a_2\propto m_f\ll 1$, where $m_b$
and $m_f$ are masses of the bosonic and fermionic components of the
$\caln=2$ hypermultiplet measured with respect to the string scale.  In
what follows, we solve for the leading order deformation in $\a_1$ and
$\a_2$.  Specifically, we seek a solution of \eqref{aeq} in the form
\begin{equation}
\begin{split}
A(y)&=\ah-\frac 14 \ln (1-y^2)+\a_1^2 A_1(y)+\a_2^2 A_2(y)\,,\\
\r(y)&=1+\a_1 \r_1(y)\,,\\
\chi(y)&=\a_2 \chi_2(y)\,.
\end{split}
\eqlabel{deformT}
\end{equation}
Substituting this ansatz into \eqref{aeq}, and working to first
non-trivial order in $\alpha_i$, we find the linearized scalar equations
\begin{equation}
\begin{split}
0&=(1-y^2)^2\,\left(y\,\r_1'\right)'+y\,\r_1\,,\\
0&=(1-y^2)^2\,\left(y\,\chi_2'\right)'+\frac34\,y\,\chi_2\,,\\
\end{split}
\eqlabel{scaleom}
\end{equation}
as well as the equations governing the back-reaction on the metric
\begin{equation}
\begin{split}
0&=y (1-y^2)\,A_1''-(1+3 y^2)\,A_1'+4 y (1-y^2)\,\left(\r_1'\right)^2\,,\\
0&=y (1-y^2)\,A_2''-(1+3 y^2)\,A_2'+\frac 43 y (1-y^2)
\,\left(\chi_2'\right)^2\,.
\end{split}
\eqlabel{defeq}
\end{equation}

The scalar equations, \eqref{scaleom}, may equally well be obtained from
the linearization of \eqref{scalar}.  Note that for an arbitrary scalar
$\Phi(y)$ of mass $m$, its equation of motion, $\jsquare\Phi=m^2\Phi$,
in the background \eqref{d3non} has the form
\begin{equation}
\Phi''+\fft1y\Phi'=\fft{(mL)^2}{4(1-y^2)^2}\Phi\,.
\end{equation}
This may be readily solved in terms of hypergeometric functions.  Although
there are generally two linearly independent solutions, only one
combination is regular at the horizon, $y\to0$.  Defining
$E_0=2\pm\sqrt{4+(mL)^2}$, the regular solution has the form
\begin{equation}
\Phi=(1-y^2)^{E_0/4}{}_2F_1(\ft14E_0,\ft14E_0,1;y^2)\,.
\eqlabel{genssoln}
\end{equation}
As a result, for the $\rho$ and $\chi$ scalars, we obtain
\begin{equation}
\begin{split}
\r_1&=(1-y^2)^{1/2}\ _2F_1(\ft12,\ft12,1; y^2)\,,\\
\chi_2&=(1-y^2)^{3/4}\ _2F_1(\ft34,\ft34,1;y^2)\,,
\end{split}
\eqlabel{answer2}
\end{equation}
where, without loss of generality, we assumed the horizon boundary conditions
\begin{equation}
\r_1\bigg|_{y=0}=1\,,\qquad \chi_2\bigg|_{y=0}=1\,.
\eqlabel{horb}
\end{equation}
Note that as $y\to 1_-$, the perturbations $\r_1,\ \chi_2$ vanish.  This
is readily seen by rewriting the solution, \eqref{genssoln}, as
\begin{equation}
\begin{split}
\Phi=&\fft{\Gamma(1-\fft12E_0)}{\Gamma(1-\fft14E_0)}(1-y^2)^{E_0/4}
{}_2F_1(\ft14E_0,\ft14E_0,\ft12E_0;1-y^2)\\
&+\fft{\Gamma(\fft12E_0-1)}{\Gamma(\fft14E_0)}(1-y^2)^{(4-E_0)/4}
{}_2F_1(1-\ft14E_0,1-\ft14E_0,2-\ft12E_0;1-y^2)\,,
\end{split}
\end{equation}
which is valid provided%
\footnote{For $\r_1$, which has $E_0=2$, the behavior as $y\to1_-$ picks
up a log, namely $\r_1\sim(1-y)^{1/2}$ and $(1-y)^{1/2}\ln(1-y)$.}
$E_0\ne2$.  Expanding for $y\to1_-$ yields the expected boundary
behavior, $\Phi\sim (1-y)^{\Delta_+/4}$ and $(1-y)^{\Delta_-/4}$,
where the conformal dimensions $\Delta_\pm$ correspond to $E_0$ and
$4-E_0$.  In this case, however, instead of being independent, the
$\Delta_+$ and $\Delta_-$ modes are related by the condition of
horizon regularity.

Turning to the leading order gravitational back-reaction, we see that
\eqref{defeq} can be solved by quadratures
\begin{equation}
\begin{split}
A_1&=\xi_1-4\int_0^y\,\frac{z\,dz}{(1-z^2)^2}\
\biggl(
\gamma_1+\int_0^z dx\,\left(\frac{\del\r_1}{\del x}
\right)^2\,
\frac{(1-x^2)^2 }{x}\biggr)\,,\\[4pt]
A_2&=\xi_2-\frac 43 \int_0^y\,\frac{z\,dz}{(1-z^2)^2}\
\biggl(
\gamma_2+\int_0^z dx\,\left(\frac{\del\chi_2}{\del x}
\right)^2\,
\frac{(1-x^2)^2 }{x}\biggr)\,,\\
\end{split}
\eqlabel{answer34}
\end{equation}
where $\gamma_i,\ \xi_i$ are four integration constants.
For a generic choice of $\gamma_i$, we find $A_i|_{y\to 1_-}
\propto (1-y)^{-1}$;
thus to recover the proper asymptotics in the UV $AdS_5\times S^5$ geometry,
these constants must be fine tuned:
\begin{equation}
\gamma_1=\frac{8-\pi^2}{2\pi^2}\,,\qquad \gamma_2=\frac{8-3\pi}{8\pi}\,.
\eqlabel{gammas}
\end{equation}
The other two integration constants, $\xi_i$, can be absorbed in a
redefinition of $\ah$. In fact, we show below that the physical quantities
are $\xi_i$ independent.  Note that at the horizon we have the behavior
\begin{equation}
\left(\frac{\del y}{\del r}\right)^2\bigg|_{y\to 0_+}
\!\!=\dd^2\ \sim\ 1+4\a_1^2 \left(
1+2 \gamma_1\right)+\a_2^2\left(1+\frac 83 \gamma_2\right)\,.
\eqlabel{dy2}
\end{equation}
This allows us to compute the BH temperature
\begin{equation}
T=\frac{1}{2\pi} e^{A} \left(\frac{\del y}{\del r}\right)\bigg|_{y\to 0_+}
\!\!=\frac{1}{2\pi} e^{\a}\,\dd
\sim\frac{1}{2\pi} e^{\ah}\,\biggl(1+\a_1^2\left(2+\xi_1+
4\gamma_1\right)+\frac 16 \a_2^2\left(3+6 \xi_2+8\gamma_2\right)\biggr).
\eqlabel{Tdef}
\end{equation}
While the first relation is valid for arbitrary temperature
and masses, in the second one we have kept only the leading
terms in $\a_i$.

Using the explicit lifting of the metric, \eqref{10m}, we may
compute the area of the BH horizon
\begin{equation}
\begin{split}
\cala_{horizon}=V_3\ e^{3 A}\bigg|_{y\to 0_+}
\!\!&=V_3\ e^{3\a}\\
&\sim V_3\ e^{3\ah}\ 2^5 {\rm vol}_{S^5}\,\left(
1+3\xi_1 \a_1^2+3\xi_2 \a_2^2\right)\,,
\end{split}
\eqlabel{ahor}
\end{equation}
where $V_3$ is the 3-dimensional volume and
${\rm vol}_{S^5}$ is the volume of the unit $S^5$.
Again, the first relation in \eqref{ahor} is exact for all
temperatures/masses.
The Bekenstein-Hawking entropy density is%
\footnote{We have used the standard relations
$16\pi G_N=(2\pi)^7 g_s^2 l_s^8$, $4\pi g_s N l_s^4=L^4$,
and the fact that we set $L=2$.}
\begin{equation}
S_{BH}=\frac {\cala_{horizon}}{4 G_N}=
\frac 12\,\pi^2 N^2\,\left(
\frac{1}{2\pi}\ e^{\a}\right)^3
\sim\frac 12\,\pi^2 N^2\,\left(
\frac{1}{2\pi}\ e^{\ah}\right)^3
\left(1+3\xi_1 \a_1^2+3\xi_2 \a_2^2\right) \,.
\eqlabel{entropy}
\end{equation}


\section{Thermodynamics and the signature of the phase transition}

In this section we discuss the thermodynamic properties of the
$N=2^*$ theory. As we have explained above, physically we expect a
phase transition between the deconfining phase (at high temperature)
and the finite temperature Coulomb phase (at low temperature).  The
high temperature phase is realized by the BH geometry, represented
by the solution to \eqref{aeq} with boundary conditions \eqref{ds0asa}.
The low temperature phase is the $\gamma=0$ (Euclidean) PW geometry
\cite{pw} with periodically identified (Euclidean) time direction
$t_E\sim t_E +1/T$.

We begin by considering the standard definition (and regularization)
of the free energy and the energy of the finite temperature deformed
PW background.  Specifically, we identify the Helmholtz free energy
$F$ with the combination $T\,I^{renom}_E$, where  $I^{renom}_E$ is
the renormalized Euclidean gravitational action, and the dual gauge
theory energy $E$ with the ADM mass of the finite temperature
deformed PW geometry. Also, we identify the gauge theory entropy
with the Bekenstein-Hawking entropy of the deformed PW background.
We show that, with such identifications, we identically satisfy the
thermodynamic relation $F=E-T S$.  This extraction of the thermodynamic
quantities are valid for arbitrary values of mass and temperature.

To proceed, we note that supersymmetry of the PW background relates
the $\a_1$ and $\a_2$ coefficients of the leading nontrivial
asymptotic behavior of the five-dimensional scalars $\r$ and $\chi$.
This allows us to parametrize the thermal phase of PW by the single
quantity $m/T$.  In the high temperature phase, we analytically
compute the leading correction to the near extremal D3 brane
thermodynamics due to the PW mass deformation.  The sign of this
correction to the black D3 branes is consistent with our claim for
a phase transition.

We find that, in the high temperature phase, our extracted free energy
no longer satisfies $dF= -S dT$, thus apparently violating the first law
of thermodynamics.  We provide a possible resolution to this puzzle in
terms of a generalized chemical potential induced by the PW mass deformation.
However a full understanding of the thermodynamics requires additional
investigation.

\subsection{The regularized free energy and the energy}

We recall that the free energy $F$, the energy $E$, and the entropy $S$
of the a system is related by the well known expression
\begin{equation}
F=E-T S \,.
\eqlabel{thermof}
\end{equation}
Here, $F$, $E$ and $S$ are well defined quantities in a weakly coupled
gauge theory, and physically should remain well defined (finite) at large
't Hooft coupling.  It is known, however, that in the dual supergravity
both the free energy and energy densities are divergent, and thus
need to be properly renormalized before they can yield a finite answer.
A standard regulating procedure is to compute such quantities by
comparison with a 'reference' supergravity background having the same
asymptotics. This comparison is often {\it ad hoc}; in particular,
the matching of the two geometries at hand at the regularization boundary
remains slightly ambiguous.  In our case, however, we have a physically
well motivated reference geometry, namely the PW background with
periodic Euclidean time direction of appropriate size.

Using the PW background as reference, strictly speaking we will not be
computing  $F_{BH}$, $E_{BH}$ and $S_{BH}$ directly [where the subscript
$ _{BH}$ relates to the non extremal deformation \eqref{ab}--\eqref{ds0asa}],
but rather the differences
\begin{equation}
\Delta F\equiv\,F_{BH}-F_{PW}\,,\qquad
\Delta E\equiv\,E_{BH}-E_{PW}\,,\qquad
\Delta S\equiv\,S_{BH}-S_{PW}\,.
\eqlabel{fes}
\end{equation}
In practice, we expect these quantities to be dominated by their
$_{BH}$ values, $\Delta F=F_{BH}$, $\Delta E=E_{BH}$, $\Delta
S=S_{BH}$.  This is clearly the case at weak 't Hooft coupling since
the thermodynamic quantities in the finite temperature Coulomb phase
are $1/N$ down compared to the corresponding quantities in the
deconfined phase, and thus the former are essentially zero in the
large $N$ limit.  Experience with other examples of the gauge/gravity
correspondence suggests that going to strong 't Hooft coupling would
typically modify the prefactor, but not the large $N$ scaling of
the free energy, the energy and the entropy.  Note that the choice of
PW background as a reference one is particularly convenient when exploring
the phase transition, as a  phase transition implies going
through a zero in $\Delta F$ in \eqref{fes} as one changes the temperature.

Before proceeding to the BH solution, we recall the asymptotics of the
reference PW geometry. In \cite{pw} the solution of the supersymmetric
$\gamma=0$ flow equations is given in terms of $\chi$ as the flow coordinate:
\begin{equation}
\begin{split}
e^A&=\frac{k \r^2}{\sinh(2\chi)}\,,\\
\r^6&=\cosh(2\chi)+\sinh^2(2\chi)\,\ln\frac{\sinh(\chi)}{\cosh(\chi)}\,.
\end{split}
\eqlabel{pwsolution}
\end{equation}
The single integration constant $k$ in \eqref{pwsolution}
is related to the hypermultiplet mass $m$ in \eqref{sp} by \cite{bpp}
\begin{equation}
k= m L =2 m\,.
\eqlabel{kim}
\end{equation}
As indicated in \eqref{pwasi} and \eqref{pwas0}, the scalars $(\rho,\chi)$
start at their UV fixed point $(1_-,0_+)$, and flow toward $(0_+,+\infty)$
in the IR.  Using the the flow equations, \eqref{pwo}, we find the IR
asymptotics
\begin{equation}
\begin{split}
\ln \r&\sim -\fft13 \chi +\fft16 \ln\fft43\,,\\
e^A&\sim 2k\,(4/3)^{1/3}\,e^{-8\chi/3}\,,\\
\frac {dA}{dr}&\sim (4/3)^{2/3}\,e^{2\chi/3}\,.
\end{split}
\eqlabel{irass}
\end{equation}
For matching, we are more interested in the UV behavior.  To develop the
asymptotics at the boundary, we introduce
\begin{equation}
\xh\equiv e^{-r/2}\,.
\eqlabel{rx}
\end{equation}
We find in the UV%
\begin{equation}
\begin{split}
\chi&\sim k\xh \left[1+ k^2\xh^2\left(\ft13+\ft{4}{3}\ln(k\xh)\right)
+k^4\xh^4\left(-\ft{7}{90}+\ft{10}{3}\ln(k\xh)+\ft{20}{9}\ln^2(k\xh)\right)
\right]\,,\\[8pt]
\r&\sim 1+ k^2\xh^2 \left(\ft13+\ft{2}{3}\ln(k\xh)\right)
+k^4\xh^4\left(\ft{1}{18}+2\ln(k\xh)+\ft23\ln^2(k\xh)\right)\,,\\[8pt]
A&\sim -\ln (2\xh)- \ft13 k^2\xh^2
-k^4\xh^4\left(\ft{2}{9}+\ft{10}{9}\ln(k\xh)+\ft{4}{9}\ln^2(k\xh)\right)
\,,\\[8pt]
\frac {dA}{dr}&\sim \ft12 + \ft13 k^2\xh^2
+k^4\xh^4\left(1+\ft{8}{3}\ln(k\xh)+\ft{8}{9}\ln^2(k\xh)\right)\,.
\end{split}
\eqlabel{uvass}
\end{equation}
Note that, as will be evident later, we need to keep terms up to
${\cal O}(\xh^4)$ in the expansion.

Turning now to the BW geometry, the general solution of \eqref{aeq}
which is smooth in the IR ($e^B\to 0$) has three integration constants,
$\{\a,\chi_0,\rho_0\}$, which are related to temperature and masses of the
$\caln=2$ hypermultiplet components, \eqref{ds0asa}, \eqref{1dsa}. The most
general solution of \eqref{aeq} in the UV ($\chi\to 0_+$) has altogether
five parameters, $\{\xi,\rh_{10},\rh_{11},\chih_0,\chih_{10}\}$.
Three of them are related to the temperature and the masses,
while the other two are uniquely determined from the requirement of having
a regular horizon, \eqref{1dsa}.  In any case, we have a three parameter
BH solution%
\begin{equation}
\begin{split}
B\sim\,&-\beta\,x^4\bigl[1+\ft89 x^2 \chih_0^2+x^4\bigl(
\ft{5}{16}\rh_{11}^2-\ft12\rh_{11}\rh_{10}+\ft{1}{18}
\chih_0^4+2\rh_{10}^2+\chih_{0}^2\chih_{10}\\
&\qquad+\ln x \left(-\ft12 \rh_{11}^2+\ft43\chih_0^4+4\rh_{11}\rh_{10}
\right)+2\rh_{11}^2 \ln^2 x\bigr)\bigr]\,,\\[8pt]
\chi\sim\,&\chih_0\,x\,\bigl[
1+x^2 \left(\chih_{10}+\ft43 \chih_0^2\ \ln x \right)
+x^4\bigl(\ft{31}{8} \rh_{11}^2-\ft{13}{2}\rh_{11}\rh_{10}
-\ft{56}{45}\chih_0^4-\ft32 \chih_0^2\rh_{11}+2\chih_0^2\rh_{10}\\
&\qquad+5 \rh_{10}^2+2\chih_0^2\chih_{10}
+\ln x\left(-\ft{13}{2}\rh_{11}^2+10\rh_{11}\rh_{10}+\ft83
\chih_0^4+2\chih_0^2\rh_{11}\right)+5\rh_{11}^2\ln^2 x\bigr)\bigr]\,,\\[8pt]
\r\sim\,&1+x^2\left(\rh_{10}+\rh_{11}\ln x\right)
+x^4\bigl(-2 \rh_{11}\rh_{10}+\ft32 \rh_{11}^2+\ft32\rh_{10}^2
+\ft{10}{3}\chih_0^2\rh_{10}-\ft{8}{3}\chih_0^2\rh_{11}+\ft13\chih_0^4\\
&\qquad+\ln x \left(3\rh_{11}\rh_{10}+\ft{10}{3}
\chih_0^2\rh_{11}-2\rh_{11}^2\right)+\ft32 \rh_{11}^2\ln^2 x\bigr)\,,\\[8pt]
A\sim\,&\xi-\ln x -\ft 13 \chih_{0}^2 x^2+x^4\bigl(\ft14\beta
+\ft19 \chih_0^4-\ft12\chih_0^2\chih_{10}-\ft{1}{8}\rh_{11}^2-\rh_{10}^2\\
&\qquad-\ln x \left(\ft23 \chih_0^4+2\rh_{11}\rh_{10}\right)
-\rh_{11}^2\ln^2 x\bigr)\,,
\end{split}
\eqlabel{uvbh1}
\end{equation}
Here we have introduced an additional integration constant $\beta$ which,
however, can be absorbed at the expense of shifting the position of the
horizon in the radial coordinate $r$ (or alternatively by rescaling $x$).
For this reason, $\beta$ should not be considered an independent parameter
of the solution.
Also, we find
\begin{equation}
\begin{split}
\frac{dA}{dr}\sim\ & \ft12 +\ft13 \chih_0^2 x^2+x^4
\bigl(-\ft12\beta +2 \rh_{10}^2+\rh_{11}\rh_{10}
+\ft14 \rh_{11}^2+\chih_0^2\chih_{10}+\ft{1}{9}\chih_{0}^4\\
&+\ln x \left(\rh_{11}^2 +\ft43 \chih_{0}^4+4\rh_{11}\rh_{10}
\right)+2\rh_{11}^2\ln^2 x\bigr)\,.
\end{split}
\eqlabel{uvbh2}
\end{equation}
In \eqref{uvbh1} and \eqref{uvbh2}, $x=x_0 e^{-r/2}$ with $x_0$ an
arbitrary constant.

The free energy, $F$, of the gravitational action can be obtained
from the (Euclidean) action  $I_E$ according to
\begin{equation}
F=T\,I_E=\frac{1}{2\pi}\,e^{\a}\,\dd\,I_E[\a,\chi_0,\rho_0]\,,
\eqlabel{fenergy}
\end{equation}
where $T$ is the temperature. As usual, $I_E$ is divergent and should be
properly regularized.  As explained above, our approach is to regulate the
free energy by subtraction, $\Delta F=T\Delta I_E$, where
\begin{equation}
\Delta I_E[\a,\chi_0,\rho_0]=\lim_{r\to\infty}\
\biggl\{I_{BH}^r[\a,\chi_0,\rho_0]-I_{PW}^r[T,k]\biggr\}\,.
\eqlabel{actreg}
\end{equation}
The regularized action $I_E^r$ consists of both volume and surface terms
\begin{equation}
\begin{split}
I_E^r=&I^r_{\rm bulk}+I^r_{\rm surf}\\
=&\frac{1}{4\pi G_5} \int^r\! dr \int_{\del \calm_5}
\!d^4\xi\,\sqrt{g^E} \left(-\ft14 R^E+(\del\a)^2+(\del\chi)^2+\calp\right)\\
&-\frac{1}{8\pi G_5}\int_{\del\calm_5}\!d^4\xi\,\sqrt{h^E}\ \nabla_\mu n^{\mu}
\,,
\end{split}
\eqlabel{regaction}
\end{equation}
where $G_5$ is the five dimensional Newton's constant
\begin{equation}
G_5\equiv \frac{G_{N}}{2^5\ {\rm vol}_{S^5}}=\frac{4\pi}{ N^2}\,,
\eqlabel{5new}
\end{equation}
$g^E$ is the Euclidean version of the metric \eqref{ab},
$n^\mu$ is a unit vector orthogonal to the four-dimensional
boundary $\del\calm_5$, and $h^E_{\mu\nu}$ is the induced
metric on $\del\calm_5$
\begin{equation}
h^E_{\mu\nu}=g^E_{\mu\nu}+n_\mu n_\nu\,.
\eqlabel{induced}
\end{equation}
In \eqref{actreg} we have assumed that the boundary $\del\calm_5$
is defined at fixed $r$ [in the coordinates \eqref{ab}],
which we will take to infinity at the end of the calculations.
In this case, the unit normal vector is $n^\mu=\delta^\mu_r$.

Consider first the bulk contribution in \eqref{regaction}.
Because of local diffeomorphism invariance, the on-shell
value of the action must reduce to a surface integral.
This is indeed what we find:%
\footnote{We omit the volume integral over the boundary $\del\calm_5$:
$\int_{\del\calm_5}d^4\xi=V_3/T$.}
\begin{equation}
\begin{split}
I^r_{\rm bulk}=&\frac{1}{4\pi G_5} \int^rdr\,\sqrt{g^E}\,
\left(-\ft23\calp\right)\\
=&\frac{1}{4\pi G_5} \int^rdr\ \left(\ft12\,e^{3 A}\,
\left(e^{A+B}\right)'+\upsilon\,e^{4 A}\left(e^B\right)'\right)'\,,
\end{split}
\eqlabel{bulk0}
\end{equation}
where $\upsilon$ is an arbitrary constant parameterizing the
constraint \eqref{intB}. In what follows, we find it convenient to
set
\begin{equation}
\upsilon=0\,.
\eqlabel{upchoice}
\end{equation}
In this case
\begin{equation}
I^r_{bulk}=\frac{1}{8\pi G_5}\,e^{3A}\left(e^{A+B}\right)'\bigg|_{horizon}^r\,,
\eqlabel{bulk}
\end{equation}
where ${horizon}$ refers to either the standard black hole horizon location
for the ``deconfining phase'' (BH) analytically continued to Euclidean
signature or the IR ($\chi\to +\infty$) of the Euclidean PW solution with
periodically identified time direction (PW).  Notice that at the black hole
horizon
\begin{equation}
\frac{1}{8\pi G_5}\, e^{3A} \left(e^{A+B}\right)'\bigg|_{horizon,\ BH}=
\frac{1}{8\pi G_5}\, e^{4A}\ \frac{\del y}{\del r}\bigg|_{horizon,\ BH}
=S_{BH} T\,,
\eqlabel{bulkhor}
\end{equation}
where $S_{BH}$ is exactly the black hole entropy \eqref{entropy}, and $T$
is the corresponding black hole temperature \eqref{Tdef}.
On the other hand, using the IR asymptotics of the PW solution, \eqref{irass},
we find instead
\begin{equation}
\frac{1}{8\pi G_5}\, e^{3A} \left(e^{A}\right)'\bigg|_{horizon,\ PW}=0
=S_{PW} T\,,
\eqlabel{bulkhorPW}
\end{equation}
which is simply interpreted in terms of the vanishing entropy
of the PW phase.

It should be noted that the black hole horizon is a regular point of the
Euclidean geometry.  Thus it is unusual to find a horizon surface term
contribution to the Euclidean bulk action \eqref{bulk0}.  In fact, this
contribution is somewhat artificial, and arises because of
our particular choice of $\upsilon$, \eqref{upchoice}.
Indeed, for generic $\upsilon$, we find
\begin{equation}
\left[\frac{1}{8\pi G_5}\, e^{3A} \left(e^{A+B}\right)'
+\frac{\upsilon}{4\pi G_5}\, e^{4 A +B} B'\right]
\bigg|_{horizon}=\frac {1 +2\upsilon}{8\pi G_5}\ e^{4 A +B} B'
\bigg|_{horizon}\,,
\eqlabel{arbup}
\end{equation}
where we have used the fact that $e^{3A+B}(e^A)'|_{horizon}=0$ for
both the $BH$ and the $PW$ phases.  {}From \eqref{arbup} we see that
for $\upsilon=-1/2$, the full contribution to $I^r_{bulk}$, \eqref{bulk0},
would come from the asymptotic region.  Although, strictly speaking, this
is the only proper value for $\upsilon$, since the full value of the
Euclidean action \eqref{bulk0} is independent of $\upsilon$, we
nevertheless find it convenient to retain $\upsilon=0$, as indicated
in \eqref{upchoice}.

For the surface term in \eqref{regaction}, we find
\begin{equation}
I^r_{surf}=-\frac{1}{8\pi G_5}\ \left(e^{4 A+B}\right)'\bigg|^r\,.
\eqlabel{surf}
\end{equation}
Adding the bulk \eqref{bulk} and the surface \eqref{surf}
terms together, we find
\begin{equation}
I_E^r=-\frac{1}{8\pi G_5} e^{3A} \left(e^{A+B}\right)'\bigg|_{horizon}
-\frac{3}{8\pi G_5} e^{4 A+B}\ A'\bigg|^r \,.
\eqlabel{total}
\end{equation}
We have shown above that the first term in \eqref{total} is simply the
combination $-S T$ where $S$ is the entropy density. If the standard
relation \eqref{thermof} is realized in the supergravity (and it must be so),
then the other term in \eqref{total} must be the regularized energy density.
Indeed this is so, provided we define the (regularized) ADM energy density
of the background as%
\footnote{As usual, the reference background has to be subtracted before
the $r\to \infty$ limit is taken.}
\begin{equation}
E\,V_3=\lim_{r\to \infty}\left\{E^r\,V_3\right\}
=\lim_{r\to \infty}\left\{-\frac{1}{8\pi G_5}\int_{v_3(r)} \sqrt{-g_{tt}}
\ {}^2 K_{v_3} \,d v_3(r)\right\}\,,
\eqlabel{energy}
\end{equation}
where the 3-boundary $v_3(r)$ is the spacelike foliation of
$\del\calm_5$ and ${}^2K_{v_3}$ is its extrinsic curvature.
Explicit evaluation of \eqref{energy} yields
\begin{equation}
E^r=-\frac{3}{8\pi G_5}\,e^{4 A+B}\ A'\bigg|^r\,,
\eqlabel{enr}
\end{equation}
which is indeed the second term of \eqref{total}.

Having derived the general asymptotic expansions for the black hole and the PW
geometry in \eqref{uvass} and \eqref{uvbh1}, we are now ready to evaluate
$\Delta F$:
\begin{equation}
\begin{split}
F_{BH}-F_{PW}=&-\left(S_{BH}-S_{PW}\right) T+ \left(E_{BH}-E_{PW}\right) \\
=& \frac{1}{8 \pi G_5}\biggl\{
-e^{4A} \frac{\del y}{\del r}\bigg|_{horizon,\ BH}-3\
\lim_{r\to \infty}\left[e^{4 A+B} A'\bigg|_{BH}-e^{4 A+B} A'\bigg|_{PW}\right]
\biggr\}\\
=& \frac{1}{8 \pi G_5}\biggl\{ -e^{4\alpha}\ \dd-3\ \Delta_{PW}^{BH}
\biggr\}\,,
\end{split}
\eqlabel{tf}
\end{equation}
where in the last line we have used \eqref{ds0asa}. The evaluation of
the limit in \eqref{tf} is rather simple.  We choose a direct matching
condition of the $BH$ and $PW$ boundaries, parameterized by $x$ [see
\eqref{uvbh1} and \eqref{uvbh2}] and $\xh$ [see \eqref{uvass}]
respectively:
\begin{equation}
\xh=\dd_0\ x\,.
\eqlabel{xxh}
\end{equation}
Additionally we have to set
\begin{equation}
B\bigg|_{PW}= 0\,.
\eqlabel{bPW}
\end{equation}
Matching the boundary values of the scalars $\r$ and $\chi$ for the black
hole and reference geometries yields
\begin{equation}
\rh_{11}=\frac 23\ k^2\ \dd_0^2,\qquad \chih_0=k\ \dd_0\,.
\eqlabel{scalm}
\end{equation}
Furthermore, matching the asymptotic volumes of the $BH$ and $PW$
phases determines
\begin{equation}
\dd_0=\frac 12 e^{-\xi}\,.
\eqlabel{dddef}
\end{equation}
The final result is
\begin{equation}
\begin{split}
\Delta^{BH}_{PW}&=e^{4\xi}\biggl(-\frac {\beta}{2}
+\frac 16\ \eta^2 \rh_{10}
-\frac{1}{72}\ \eta^4 \ln \left(\frac 14\ e\ \eta^2\right)\biggr)\,,
\end{split}
\eqlabel{finalans}
\end{equation}
where we have introduced
\begin{equation}
\eta\equiv k e^{-\xi}\,.
\eqlabel{edef}
\end{equation}

The difference of free energies thus has the form
\begin{equation}
\begin{split}
F_{BH}-F_{PW}&=-\frac{e^{4\a}}{8\pi G_5}\ \biggl(\dd+3\,e^{-4\a}
\Delta^{BH}_{PW}\biggr)\\
&=-\frac{\pi^2 N^2}{2}\ \left(\frac{1}{2\pi}\ e^{\a}\right)^4
\biggl(\dd+3\ e^{-4\a} \Delta^{BH}_{PW}\biggr)\,.
\end{split}
\eqlabel{finaldF0}
\end{equation}
This can be further simplified by using the integral of motion, \eqref{intB}.
Indeed, evaluating the constant in \eqref{intB} in the IR and the UV, and
equating them, we find
\begin{equation}
\ln\beta+\ln 2+4\xi=4\a+\ln\dd\,.
\eqlabel{relxial}
\end{equation}
Thus we can rewrite \eqref{finaldF0} as
\begin{equation}
\begin{split}
F_{BH}-F_{PW}=-\frac{\pi^2 N^2}{4 (2\pi)^4}\,e^{4\xi}\,
\biggl(\beta+\eta^2 \rh_{10} -\frac {1}{12}\,\eta^4\,\ln\left(
\frac 14 \,e\,\eta^2\right)\biggr)\,.
\end{split}
\eqlabel{finaldF}
\end{equation}
Notice that from \eqref{uvbh1} the residual reparametrization invariance
$x\to \lambda x$ can be absorbed by the following transformation on the
quantities $\{\xi,\rh_{10},\rh_{11},\chih_0,\chih_{10},\beta\}$:
\begin{equation}
\begin{split}
\xi&\to \xi-\ln \lambda\,,\\
\rh_{10}&\to \lambda^2 \rh_{10}+\lambda^2 \rh_{11}\ln \lambda\,,\qquad
\rh_{11}\to \lambda^2 \rh_{11}\,,\\
\chih_0&\to \lambda \chih_0\,,\kern7.8em
\chih_{10}\to \lambda^2 \chih_{10}+\frac 43 \lambda^2 \chih_0^2\ln \lambda\,,\\
\beta&\to \lambda^4 \beta\,.
\end{split}
\eqlabel{transf}
\end{equation}
This leaves \eqref{finaldF} invariant.

{}From the gauge theory arguments, we expect that the free energy of
the $PW$ phase scales as $N^{1}$.  On the other hand, the $N$-scaling in
\eqref{finaldF} suggests that in the large $N$-limit, $F_{PW}$ is essentially
zero compared to $F_{BH}$. Thus, in the high temperature phase, we identify
the $\caln=2^*$ gauge theory Helmholtz free energy density $F$ at large
't Hooft coupling with \eqref{finaldF}
\begin{equation}
\begin{split}
F\equiv F_{BH}-F_{PW}&\equiv T\, I_E^{renom}\\
&=-\frac{\pi^2 N^2}{4 (2\pi)^4}\,e^{4\xi}\,
\biggl(\beta+\eta^2 \rh_{10} -\frac {1}{12}\,\eta^4\,\ln\left(
\frac 14\,e\,\eta^2\right)\biggr)\,.
\end{split}
\eqlabel{fgauge}
\end{equation}
Also, from \eqref{entropy} and \eqref{relxial}, the $\caln=2^*$ gauge
theory entropy density $S$ is
\begin{equation}
S\equiv S_{BH}=\frac{N^2}{8 (2\pi)}\,e^{3\a}=
\frac{N^2}{8 (2\pi)}\,\left(\frac{2\beta}{\delta}\right)^{3/4}\,e^{3\xi}\,.
\eqlabel{sgauge}
\end{equation}
Finally, the gauge theory energy density $E$ is that of the
(renormalized) ADM energy density [from \eqref{tf}]
\begin{equation}
E\equiv E_{BH}-E_{PW}=T\, I_E^{renom}+T\, S_{BH}\,.
\eqlabel{egauge}
\end{equation}
The gauge theory temperature $T$ is identified with that of the horizon
in the $BH$ phase, \eqref{Tdef}.

\subsection{The high temperature thermodynamics of the $\caln=2^*$}

Given the analytical expression for the high temperature expansion,
\eqref{deformT}--\eqref{gammas}, it is straightforward to determine the
leading correction to the non-extremal D3 brane thermodynamics due to the
PW mass flow. As we will note, the sign of this correction suggests the
possibility of the ``deconfinement $\rightarrow$ finite temperature
Coulomb phase'' phase transition.

We begin by matching the $\{\ah,\a_1,\a_2\}$ parameters in \eqref{deformT}
with $\{\xi,\chih,\rh\}$ of the asymptotic expansion, \eqref{uvbh1}.
Recall that the $y$ coordinate in \eqref{answer2} is just $e^B$.  Thus
the $x$ coordinate of \eqref{uvbh1} and $y$ are related according to
\begin{equation}
y\sim\ 1-2 x^4\,.
\eqlabel{xy}
\end{equation}
This corresponds to setting $\beta=2$ in \eqref{uvbh1}. Now, to linear
order in $\a_i$, by matching the scalars $\r$, $\chi$ in \eqref{answer2}
and \eqref{uvbh1}, we find
\begin{equation}
\begin{split}
\rh_{10}&=\frac{4\ln 2}{\pi}\,\a_1\,,
\qquad\qquad\quad\,\rh_{11}=-\frac{8}{\pi}\,\a_1\,,\\
\chih_0&=\frac{\sqrt{2\pi}}{[\Gamma \left(\frac 34\right)]^2}\,\a_2\,,
\qquad\qquad\chih_{10}=-\frac{2\,[\Gamma\left(\frac 34 \right)]^4}{\pi^2}\,.
\end{split}
\eqlabel{scmatch}
\end{equation}
Notice that $\chih_{10}$ is independent of $\a_2$. Furthermore, matching to
the asymptotic PW solution, \eqref{scalm} and \eqref{dddef}, determines
\begin{equation}
\a_1=-\frac{\pi}{48}\,k^2\,e^{-2\xi}\,,\qquad\qquad
\a_2=\frac{[\Gamma\left(\frac34\right)]^2}{2^{3/2}\sqrt{\pi}}\,k\,e^{-\xi}\,.
\eqlabel{aifind}
\end{equation}
Notice that, to leading order, $\a_1\propto \a_2^2$.  This is consistent
with the gauge theory expectation that, to leading order in the $\caln=2$
hypermultiplet mass, it is enough to turn on only the mass for the fermionic
components. Thus the consistency of the high temperature
expansion (conditional to the asymptotic $\caln=2$ supersymmetry)
requires setting $\a_1$ to zero.

Now, matching $A$, we find
\begin{equation}
0=\ah-\ft12\ln 2+\a_2^2\,\hat{\xi}_2-\xi,\qquad \a=\ah+\alpha_2^2\,\xi_2\,.
\eqlabel{amatch}
\end{equation}
where [using \eqref{answer34}]
\begin{equation}
\begin{split}
\hat{\xi}_2=A_2\bigg|_{y\to 1_-}&=
\xi_2+\frac 43 \int_0^1\,\frac{z\ dz}{(1-z^2)^2}\,
\biggl(\int_z^1\,dx\,\left(\frac{\del\chi_2}{\del x}
\right)^2\,\frac{(1-x^2)^2 }{x}\biggr)\,.
\end{split}
\eqlabel{xi2h}
\end{equation}
with $\chi_2$ given by \eqref{answer2}.
There is a nontrivial check on the computation.
With \eqref{amatch} and \eqref{dy2}, we find from \eqref{relxial}
\begin{equation}
\left(\ln2\right)
+\ln 2+ 4\left(\hat{\alpha}-\ft12 \ln 2+\alpha_2^2\ \hat{\xi}_2\right)
=4\left(\ah+\a_2^2\ \xi_2\right)+\a_2^2 \left(\ft12+
\ft86\gamma_2\right)\,,
\eqlabel{consist1}
\end{equation}
or
\begin{equation}
4\left(\hat{\xi}_2-\xi_2\right)=\ft12 +\ft83 \gamma_2
=4\,\frac {1}{3\pi}\,.
\eqlabel{consist2}
\end{equation}
Given expression \eqref{xi2h}, we have numerically verified that
\eqref{consist2} is indeed correct.

Using \eqref{Tdef}, \eqref{fgauge} and \eqref{sgauge}, we can now
express the gauge theory thermodynamic quantities in the high
temperature regime in terms of $\{\xi,\eta\equiv k e^{-\xi}\ll 1\}$:
\begin{equation}
\begin{split}
T&=\frac{1}{2^{1/2}\pi}\,e^{\xi}\,\left(1+\frac {\Gamma(3/4)^4}{8\pi^2}\,
\eta^2\right)\,,\\
S&=\frac{N^2}{2^{5/2}\pi}\,e^{3 \xi}\,\left(1-\frac{\Gamma(3/4)^4}{8\pi^2}\,
\eta^2\right)\,,\\
F=E-S T&=-\frac{N^2}{32\pi^2}\,e^{4 \xi}\,.
\end{split}
\eqlabel{defd3b}
\end{equation}
Recalling \eqref{kim}, and inverting the $T\leftrightarrow \xi$ relation above
\begin{equation}
e^{\xi}=2^{1/2}\pi T \left(1-\frac {\Gamma(3/4)^4}{4\pi^4}\ \frac{m^2}{T^2}
\right)\,,
\eqlabel{xiT}
\end{equation}
we finally obtain the thermodynamic quantities
\begin{equation}
\begin{split}
S&=\frac 12 \pi^2 N^2 T^3   \left(1-
\frac{\Gamma(3/4)^4}{\pi^4}\,\frac{m^2}{T^2}\right)\,,\\
E&=\frac 38 \pi^2 N^2 T^4 \left(1-
\frac{\Gamma(3/4)^4}{\pi^4}\,\frac{m^2}{T^2}\right)\,,\\
F&=-\frac 18 \pi^2 N^2 T^4  \left(1-
\frac{\Gamma(3/4)^4}{\pi^4}\,\frac{m^2}{T^2}\right)\,.
\end{split}
\eqlabel{thf0}
\end{equation}
For the thermodynamic process \eqref{thf0} we
find%
\footnote{We have confirmed the leading correction to the free energy
in \eqref{thf0}, and thus the violation of the first law of thermodynamics,
numerically. For details see section 6.4.}
\begin{equation}
T\ dS\ne d E\,.
\eqlabel{dsde}
\end{equation}
In the next subsection we discuss a possible resolution of this
puzzle, \eqref{dsde}.

\subsection{The chemical potential of the $\caln=2^*$ flow?}

We suggest here%
\footnote{We would like to thank 
 Chris  Herzog,
David Lowe and Andrei Starinets for very
useful discussions.}
that the apparent violation of the first law of thermodynamics for the
leading in $m/T\ll 1$ correction to the high temperature thermodynamics
of the $\caln=2^*$ gauge theory, \eqref{dsde}, could be explained as due
to the neglection of the induced chemical potential for the temperature
deformed $\caln=2^*$ flow.  We stress that, while a certain chemical
potential appears to resolve the paradox, we do not have an understanding
of what exactly is its corresponding conjugate operator.  Additionally, it
is conceivable that a different subtraction procedure for the computation
of the supergravity effective action and the ADM mass would resolve the
problem with the first law of thermodynamics altogether \cite{wip}. Having
said this, however, here we restrict our attention to the possibility of
having an induced chemical potential for the temperature deformed PW flow.

One of the basic statements of the gauge/string
theory correspondence \cite{a9905} is the identification of the
type IIB string theory partition function with the $\caln=4$
gauge theory partition function, where the boundary values of the
string fields $\Phi$ are the sources of the gauge theory operators
$\calo^{\Phi}$
\begin{equation}
Z_{string}[\Phi]\bigg|_{\Phi(r\to\infty)=\Phi_0}\equiv Z_{gauge}[\Phi_0]
=e^{-W_{gauge}[\Phi_0]}\,,
\eqlabel{partf}
\end{equation}
where $W_{gauge}$ is the generating functional for the connected
Green's function in the gauge theory
\begin{equation}
W_{gauge}[\Phi_0]=-\ln \bigg\langle e^{\int d^4 x\ \Phi_0\calo^{\Phi}}
\bigg\rangle_{gauge}\,.
\eqlabel{w}
\end{equation}
It is not known how to precisely define the string theory partition
function. But, ignoring all the stringy $\a'$ corrections%
\footnote{Neglecting $\a'$ corrections implies that the string fields
$\Phi$ must actually be type IIB supergravity modes.},
and also all the string loop corrections (which basically amounts to
taking the $N\to\infty$ limit with the large but finite 't Hooft coupling),
it is reasonable to assume that $Z_{string}$ is dominated by its saddle
point%
\footnote{The subtleties of multiple saddle points will not arise
in the present situation, namely the high temperature phase of the string
theory dual to $\caln=2^*$ gauge theory.}
--- the extremum of the (Euclidean) supergravity action $I_E$ with the
prescribed boundary values of the sources $\Phi_0$:
\begin{equation}
-\ln \left(Z_{string}[\Phi]\bigg|_{\Phi(r\to\infty)=\Phi_0}\right)
\simeq
\ {\rm extremum}\ I_E[\Phi]\bigg|_{\Phi(r\to\infty)=\Phi_0}\,.
\eqlabel{isugra}
\end{equation}

We restrict to the $\caln=4$ gauge theory deformations which
are irrelevant in the UV (the finite temperature $\caln=2^*$
flow discussed in previous sections is precisely of this type).
This implies that the asymptotic geometry that extremizes
$I_E$ is necessarily $AdS_5\times S^5$
\begin{equation}
ds_{10}^2\bigg|_{r\to \infty}\simeq e^{-2r/L} dx_4^2 +dr^2+ L^2
d\Omega_5^2\,.
\eqlabel{asmetric}
\end{equation}
Generically, the supergravity mode $\Phi$ that extremizes $I_E$ behaves as
\begin{equation}
\Phi\bigg|_{r\to\infty}\sim \Phi_0\ e^{(\Delta-4)r/L}
+\calf_0\ e^{-\Delta r/L}\,,
\eqlabel{assphi}
\end{equation}
where $\Delta$ is the mass dimension of the gauge theory operator
$\calo^{\Phi}$, and $\calf_0$ should be interpreted as its
vacuum expectation value:
\begin{equation}
\langle 0_H|\calo^{\Phi}|0_H\rangle=\calf_0\,,
\eqlabel{vev}
\end{equation}
where $|0_H\rangle$ is a vacuum state of the deformed $\caln=4$
Hamiltonian $H$
\begin{equation}
H=H_{\caln=4}+\Phi_0\calo^{\Phi}\,.
\eqlabel{h}
\end{equation}

As was emphasized in \cite{ps}, in a theory with a unique
(or at least isolated) vacuum, the dynamics should determine the
vev \eqref{vev} once the Hamiltonian \eqref{h} is specified.
Generically we expect an isolated vacuum whenever the
$\caln=4$ supersymmetry is completely broken. This will always
be the case whenever arbitrary deformations of the type
\eqref{h} are supplemented by the finite temperature deformation.
Consider now such a deformation, namely gauge theory with
Hamiltonian \eqref{h} at finite temperature.
{}From the gauge theory perspective we can definite
two different partition functions:
a canonical partition function\footnote{We assume that the
gauge theory volume is constant.}
\begin{equation}
Z_{gauge}[T]=e^{-\frac 1T F[T]}=\tr e^{-\frac 1T H}\,,
\eqlabel{can}
\end{equation}
where the $\tr$ is taken over the eigenstates of the full Hamiltonian,
or the grand canonical partition function
\begin{equation}
\Xi_{gauge}[T,\mu_{\Phi}]=e^{-\frac 1T \Omega[T,\mu_{\Phi}]}=
\tr e^{-\frac 1T H +\mu_{\Phi} Q_{\Phi} }\,.
\eqlabel{grandcan}
\end{equation}
There the trace is taken over the eigenstates of both the Hamiltonian
$H$ and the (conserved) ``charge'' operator $Q_{\Phi}$, conjugate to
the chemical potential
\begin{equation}
\mu_{\Phi}=-\frac {\Phi_0}{T}\,.
\eqlabel{chempot}
\end{equation}

When we neglect the fluctuations in $Q_{\Phi}$, we obtain
\begin{equation}
\begin{split}
\Omega[T,\mu_{\Phi}]=& - T\ \ln\ \Xi_{gauge}[T,\mu_{\Phi}]\\
& \simeq F[T]-\mu_{\Phi} \langle Q_{\Phi}\rangle V_3\,,
\end{split}
\eqlabel{nofluc}
\end{equation}
where $V_3$ is the spatial volume of the gauge theory coming from the
integration over the zero momentum modes.
If we identify the string theory partition function \eqref{isugra} with the
canonical partition function of the gauge theory \eqref{can},
or equivalently
\begin{equation}
F[T]=T \ln \bigg\langle e^{\frac 1T \int d^3x\ \Phi_0\calo}
\bigg\rangle_{gauge}
\simeq  T\ {\rm extremum}\ I_E[\Phi]\bigg|_{\Phi(r\to\infty)=\Phi_0}\,,
\eqlabel{scan}
\end{equation}
in the case of the finite temperature $\caln=2^*$ gauge/string duality
we will face the breakdown of the first law of thermodynamics,
\eqref{dsde}.
Rather, we propose that one should identify
the string theory partition function \eqref{isugra} with the
{\it grand canonical} partition function of the gauge theory \eqref{nofluc},
or equivalently,
\begin{equation}
\begin{split}
\Omega[T,\mu_\Phi]&=T
\ln \bigg\langle e^{\frac 1T \int d^3x\ \Phi_0\calo}\bigg\rangle_{gauge}=T
\ln \bigg\langle e^{-\frac 1T \int d^3x\ \mu_\Phi Q_{\Phi}}
\bigg\rangle_{gauge}\\
&\simeq F[T]-\mu_{\Phi} \langle Q_{\Phi}\rangle V_3\\
&\simeq  T\ {\rm extremum}\ I_E[\Phi]\bigg|_{\Phi(r\to\infty)=\Phi_0}
\equiv T\ I_E[\Phi_0]\,,
\end{split}
\eqlabel{sgcan}
\end{equation}
where the last equivalence defines $I_E[\Phi_0]$. Notice that
the one-point correlation function  $\langle Q_{\Phi}\rangle$
can be computed by differentiating with respect to $\mu_\Phi$
the correspondence \eqref{sgcan}:
\begin{equation}
\frac{\del \Omega[T,\mu_\Phi]}{\del \mu_\Phi}
=T\left(\frac {-1}{T}\int d^3 x \langle Q_{\Phi}\rangle\right)
=-V_3 \langle Q_{\Phi}\rangle\,.
\eqlabel{veve}
\end{equation}
{}From \eqref{sgcan} and \eqref{veve} we find
\begin{equation}
F[T]=\Omega[T,\mu_\Phi]-\mu_\Phi \frac{\del \Omega[T,\mu_\Phi]}{\del
\mu_\Phi}\,.
\eqlabel{finalf}
\end{equation}
Finally, the first law of thermodynamics (for $dV_3$=0)
takes the form
\begin{equation}
dF = -S\ dT  +\mu_\Phi\ d\left(\langle Q_{\Phi}\rangle\right)\,,
\eqlabel{1stlaw}
\end{equation}
where the independent variables are $\mu_\Phi$ and $T$.

In the rest of this subsection we demonstrate that with
interpretation \eqref{sgcan}, there is no conflict with the
high temperature thermodynamics of the $\caln=2^*$ flow.
First of all, notice that though the $\caln=2^*$ flow
necessarily has moduli, its finite temperature deformation
should not. Thus we expect that turning on
the mass term for the fermions (supergravity dual to the
five-dimensional scalar $\chi$) should uniquely fix their condensate.
In the language of the asymptotic behavior of the scalar $\chi$
in \eqref{uvbh1}, this implies that specifying $\chih_0$ should
uniquely determine the coefficient of its normalizable mode
$\propto \chih_0 \chih_{10}$.  This is  precisely what we find
in \eqref{scmatch}.  If in our case we take $\Phi_0\equiv m$, we would
obtain $\mu\equiv -m/T$.  Following \eqref{sgcan}%
\footnote{As before we talk about densities of the thermodynamic quantities.},
\begin{equation}
\Omega[T,\mu]=T I_E[m]=-\frac {\pi^2 N^2 T^4}{8}+\frac {\kappa}{8}\ m^2 T^2
=-\frac {\pi^2 N^2 T^4}{8}+\frac {\kappa}{8}\ \mu^2 T^4\,,
\eqlabel{redw}
\end{equation}
where in the second equality we have substituted the Helmholtz free energy
from \eqref{thf0}, which  {\it by computation} equals $T\, I_E[m]$.
Additionally, to avoid cluttering the formulas we set
\begin{equation}
\kappa\equiv \frac {N^2}{\pi^2}\  \Gamma(3/4)^4\,.
\eqlabel{kappa}
\end{equation}
Notice that the entropy is \eqref{thf0}
\begin{equation}
S=\frac{\pi^2 N^2 T^3}{2}-\frac {\kappa}{2}\ \mu^2 T^3\,.
\end{equation}
{}From \eqref{veve} and \eqref{finalf}, we find
\begin{equation}
\begin{split}
\langle Q_\mu\rangle&=-\frac{\kappa}{4} \mu T^4\,,\\
F&=-\frac {\pi^2 N^2 T^4}{8}-\frac {\kappa}{8}\ \mu^2 T^4\,.
\end{split}
\eqlabel{2last}
\end{equation}
It is easy to see that the first law of thermodynamics, \eqref{1stlaw},
is now satisfied:
\begin{equation}
\begin{split}
dF&=\left(-\frac{\pi^2 N^2 T^3}{2}-\frac{\kappa\mu^2 T^3}{2}\right)\ dT
-\frac{\kappa \mu T^4}{4}\ d\mu\\
&=-\left(\frac{\pi^2 N^2 T^3}{2}-\frac {\kappa}{2}\ \mu^2 T^3
\right)\ dT+\mu\ d\left(-\frac{\kappa\mu T^4}{4}\right)\\
&\equiv -S\ dT+\mu\ d\left(\langle Q_\mu\rangle\right)\,.
\end{split}
\eqlabel{1last}
\end{equation}

Thus we have shown that if we assume that the finite temperature
deformation of the PW flow has an induced chemical potential
$\mu\equiv -m/T$, the interpretation of the supergravity computation
in terms of the grand canonical ensemble appears to resolve the puzzle
with the first law, \eqref{dsde}.  What is not clear, however, is what
is exactly the charge operator $Q_\mu$ conjugate to $\mu$.
Though it appears that the expectation value of $Q_\mu$, \eqref{2last},
is related to the gaugino condensate, $Q_\mu$ cannot be
the fermion mass operator; the latter does not commute with the
gauge theory Hamiltonian and thus cannot be conserved.

\subsection{The phase transition}

Independent of the high temperature expansion, the general expression
for the generalized free energy density difference between the $BH$ and
the $PW$ phases is given by \eqref{finaldF}:
\begin{equation}
\delta\Omega[T,\mu]\equiv
\Omega_{BH}-\Omega_{PW}=-\frac{\pi^2 N^2}{4 (2\pi)^4}\,e^{4\xi}\,
\biggl(\beta+\eta^2 \rh_{10} -\frac {1}{12}\ \eta^4\ \ln\left(
\frac 14 \ e\ \eta^2\right)
\biggr)\,.
\eqlabel{domega}
\end{equation}
Following the discussion of the previous subsection, we have
reinterpreted the Helmholtz free energy $F$ as the generalized
free energy $\Omega[T,\mu]$.  This expression is valid for arbitrary
temperature $T$ and chemical potential $\mu=-m/T$, which in turn are
implicitly related to the parameters of the supergravity solution,
$\xi$, $\beta$, $\eta$ and $\hat\rho_{10}$, which show up on the right hand
side of \eqref{domega}.  

To proceed beyond the high temperature expansion, we may examine
the behavior of $\delta\Omega$ numerically.  To do so, we extract
the appropriate coefficients governing the behavior of $\delta\Omega$
by matching the UV behavior of the numerical solution with \eqref{uvbh1}
and the IR behavior with \eqref{ds0asa}.  In particular, we first work
in the UV and fix the $x$-coordinate of \eqref{uvbh1} through the
functional dependence of the scalar $B$ (recalling that $\beta$ may be
scaled away).  Then, after matching the coefficients of the leading
nontrivial asymptotics $\{\rh_{11},\chih_{0}\}$ with the asymptotic PW
geometry according to \eqref{scalm} and \eqref{dddef}, we may
unambiguously extract the subleading terms $\{\rh_{10}, \chih_{10}\}$.
We finally obtain $\xi$ through the relation \eqref{relxial}, where $\a$
and $\dd$ are determined from the behavior of $A$ and $B$ at the horizon.

Of course, $\{\xi,\rh_{10},\chih_{10}\}$ are functions of the data at
the horizon $\{\a,\r_0,\chi_0\}$, \eqref{ds0asa}.  Actually $\r_0$ and
$\chi_0$ cannot be independent since the coefficients of the leading UV
asymptotics of $\r$ and $\chi$, $\{\rh_{11},\chih_{0}\}$, must satisfy
\eqref{scalm}
\begin{equation}
\frac{\rh_{11}}{\chih_{0}^2}=\frac 23\,,
\eqlabel{rationr11}
\end{equation}
which is just the statement of asymptotic $\caln=2$ supersymmetry.  This
results in a reduction to two parameters, $T$ and $\mu$ (or equivalently
$T$ and $m$).  Finally, since any scale in the $\caln=2^*$ theory may be
related to $m$, we note that, for the numerical work, we only need to
examine a one parameter set of solutions.

\begin{figure}
\begin{center}
\epsfig{file=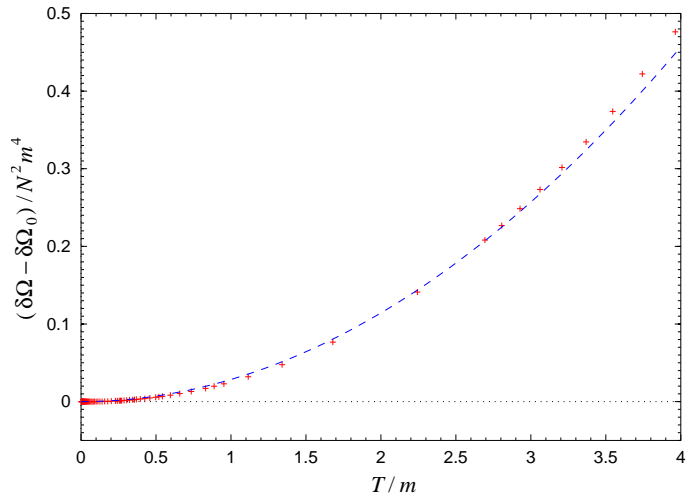,width=0.6\textwidth}
\caption{Numerical determination of the generalized 
free energy $\delta\Omega$ \eqref{domega} with the leading
behavior $\delta\Omega_0=-\pi^2N^2T^4/8$ subtracted.  
The numerical values are
given by the data points, while the leading high-temperature correction,
$\delta\Omega_1\equiv\Gamma(3/4)^4N^2m^2T^2/8\pi^2$, 
is indicated by the dashed line.}
\label{fig:freee}
\end{center}
\end{figure}

For a thermodynamic process at a fixed volume, the physical phase is
realized from the minimization of the generalized free energy
\begin{equation}
\Omega[T,\mu]^{physical}=\min \left\{\Omega_{BH},
\Omega_{PW}\right\}\,.
\end{equation}
Thus the signature of a phase transition would be the vanishing of
$\delta\Omega$ at a certain critical temperature $T_c$
\begin{equation}
\delta\Omega\left[T,\mu=-m/T\right]\bigg|_{T=T_c}=0\,.
\end{equation} 
In the high temperature phase, $m/T \ll 1$, we have found 
[compare with \eqref{redw}]
\begin{equation}
\begin{split}
\frac{\delta\Omega\left[T,-\frac mT\right]}{m^4 N^2}
\equiv& \frac{1}{m^4 N^2} \biggl(\delta\Omega_0+\delta\Omega_1\biggr)+
o\left(\frac {T^2}{m^2}\right)\\
=& -\frac {\pi^2}{8} \left(\frac Tm\right)^4
+\frac{\Gamma(3/4)^4}{8 \pi^2} \left(\frac Tm\right)^2
+o\left(\frac {T^2}{m^2}\right)\,;
\end{split}
\eqlabel{anpred}
\end{equation}
that is, $\delta\Omega <0$. On the other hand, we have argued 
that in the low temperature phase, $m/T \gg 1$,
we would instead expect $\delta\Omega >0$; see Fig.~\ref{cases}.

The result of the numerical work is shown in Fig.~\ref{fig:freee}.
While the numerics appear to be in good agreement%
\footnote{Notice that this is a numerical confirmation of the 
paradox with the first law of thermodynamics \eqref{dsde},
further discussed in section 6.3.}
with our analytical prediction \eqref{anpred}, we have been unable to
confirm the phase transition.  One possibility is that the critical
temperature of the conjectured phase transition is at $T_c=m\,\varrho$,
where $\varrho$ is a small number.  This would make numerical study of the
transition rather challenging, as the ultra-low temperature supergravity
flows (see Fig.~\ref{fig:flow}) approach the supersymmetric (singular) PW
flow, and are plagued by numerical instabilities.  Another possibility is
that in the large $N$ limit this phase transition is actually at $T_c=0$.
This issue clearly deserves further investigation.


\section*{Acknowledgments}

We would like  to thank Ofer Aharony, Martin Kruczenski, Finn Larsen,
David Lowe, Bob McNees, Rob Myers, Radu Roiban, Andrei Starinets,
Arkady Tseytlin and Arkady Vainshtein for useful discussions. 
We are especially grateful to Chris Herzog for many illuminating 
discussions and for important comments on the draft.
A.B.~would like to thank the organizers of the
``QCD and String Theory'' workshop at INT, University of Washington, for
providing an inspiring environment.  This work was supported in part by
the US Department of Energy under grant DE-FG02-95ER40899.



\begin{thebibliography}{99}

\bibitem{m9711}
J.~M.~Maldacena,
``The large $N$ limit of superconformal field theories and supergravity,''
Adv.\ Theor.\ Math.\ Phys.\  {\bf 2}, 231 (1998)
[Int.\ J.\ Theor.\ Phys.\  {\bf 38}, 1113 (1999)]
[arXiv:hep-th/9711200].

\bibitem{a9905}
O.~Aharony, S.~S.~Gubser, J.~M.~Maldacena, H.~Ooguri and Y.~Oz,
``Large $N$ field theories, string theory and gravity,''
Phys.\ Rept.\  {\bf 323}, 183 (2000)
[arXiv:hep-th/9905111].

\bibitem{w97}
E.~Witten,
``Anti-de Sitter space, thermal phase transition, and confinement in
gauge theories,''
Adv.\ Theor.\ Math.\ Phys.\  {\bf 2}, 505 (1998)
[arXiv:hep-th/9803131].

\bibitem{b0011}
A.~Buchel,
``Finite temperature resolution of the Klebanov-Tseytlin singularity,''
Nucl.\ Phys.\ B {\bf 600}, 219 (2001)
[arXiv:hep-th/0011146].

\bibitem{ghkt}
S.~S.~Gubser, C.~P.~Herzog, I.~R.~Klebanov and A.~A.~Tseytlin,
``Restoration of chiral symmetry: A supergravity perspective,''
JHEP {\bf 0105} (2001) 028
[arXiv:hep-th/0102172].

\bibitem{bf}
A.~Buchel and A.~R.~Frey,
``Comments on supergravity dual of pure $N = 1$ super Yang Mills theory
with unbroken chiral symmetry,''
Phys.\ Rev.\ D {\bf 64}, 064007 (2001)
[arXiv:hep-th/0103022].

\bibitem{gtv}
S.~S.~Gubser, A.~A.~Tseytlin and M.~S.~Volkov,
``Non-Abelian 4-d black holes, wrapped 5-branes, and their dual
descriptions,''
JHEP {\bf 0109}, 017 (2001)
[arXiv:hep-th/0108205].

\bibitem{hp}
S.~W.~Hawking and D.~N.~Page,
``Thermodynamics Of Black Holes In Anti-de Sitter Space,''
Commun.\ Math.\ Phys.\  {\bf 87},

\bibitem{kt}
I.~R.~Klebanov and A.~A.~Tseytlin,
``Gravity duals of supersymmetric $SU(N) \times SU(N+M)$ gauge theories,''
Nucl.\ Phys.\ B {\bf 578}, 123 (2000)
[arXiv:hep-th/0002159].

\bibitem{ks}
I.~R.~Klebanov and M.~J.~Strassler,
``Supergravity and a confining gauge theory: Duality cascades and
chiSB-resolution of naked singularities,''
JHEP {\bf 0008}, 052 (2000)
[arXiv:hep-th/0007191].

\bibitem{mn}
J.~M.~Maldacena and C.~Nunez,
``Towards the large $N$ limit of pure $N = 1$ super Yang Mills,''
Phys.\ Rev.\ Lett.\  {\bf 86}, 588 (2001)
[arXiv:hep-th/0008001].

\bibitem{lst}
A.~Buchel,
``On the thermodynamic instability of LST,''
arXiv:hep-th/0107102.

\bibitem{pw}
K.~Pilch and N.~P.~Warner,
``$N = 2$ supersymmetric RG flows and the IIB dilaton,''
Nucl.\ Phys.\ B {\bf 594}, 209 (2001)
[arXiv:hep-th/0004063].

\bibitem{bpp}
A.~Buchel, A.~W.~Peet and J.~Polchinski,
``Gauge dual and noncommutative extension of an $N = 2$ supergravity
solution,''
Phys.\ Rev.\ D {\bf 63}, 044009 (2001)
[arXiv:hep-th/0008076].

\bibitem{cl}
N.~Evans, C.~V.~Johnson and M.~Petrini,
``The enhancon and $N = 2$ gauge theory/gravity RG flows,''
JHEP {\bf 0010}, 022 (2000)
[arXiv:hep-th/0008081].

\bibitem{phil}
R.~Donagi and E.~Witten,
``Supersymmetric Yang-Mills Theory And Integrable Systems,''
Nucl.\ Phys.\ B {\bf 460}, 299 (1996)
[arXiv:hep-th/9510101].

\bibitem{rks}
C.~P.~Herzog, I.~R.~Klebanov and P.~Ouyang,
``D-branes on the conifold and $N = 1$ gauge / gravity dualities,''
arXiv:hep-th/0205100.

\bibitem{fm}
D.~Z.~Freedman and J.~A.~Minahan,
``Finite temperature effects in the supergravity dual of the $N = 1^*$ gauge
theory,''
JHEP {\bf 0101}, 036 (2001)
[arXiv:hep-th/0007250].

\bibitem{ps}
J.~Polchinski and M.~J.~Strassler,
``The string dual of a confining four-dimensional gauge theory,''
arXiv:hep-th/0003136.

\bibitem{HH}
S.~W.~Hawking and G.~T.~Horowitz,
``The Gravitational Hamiltonian, action, entropy and surface terms,''
Class.\ Quant.\ Grav.\  {\bf 13}, 1487 (1996)
[arXiv:gr-qc/9501014].

\bibitem{sk1}
M.~Bianchi, D.~Z.~Freedman and K.~Skenderis,
``How to go with an RG flow,''
JHEP {\bf 0108}, 041 (2001)
[arXiv:hep-th/0105276].

\bibitem{sk2}
M.~Bianchi, D.~Z.~Freedman and K.~Skenderis,
``Holographic renormalization,''
Nucl.\ Phys.\ B {\bf 631}, 159 (2002)
[arXiv:hep-th/0112119].

\bibitem{st}
G.~Policastro, D.~T.~Son and A.~O.~Starinets,
``From AdS/CFT correspondence to hydrodynamics,''
JHEP {\bf 0209}, 043 (2002)
[arXiv:hep-th/0205052].

\bibitem{b}
A.~Buchel,
``Comments on fractional instantons in $N = 2$ gauge theories,''
Phys.\ Lett.\ B {\bf 514}, 417 (2001)
[arXiv:hep-th/0101056].

\bibitem{gkp}
S.~S.~Gubser, I.~R.~Klebanov and A.~W.~Peet,
``Entropy and Temperature of Black 3-Branes,''
Phys.\ Rev.\ D {\bf 54}, 3915 (1996)
[arXiv:hep-th/9602135].

\bibitem{schwarz83} J.~H.~Schwarz,
``Covariant Field Equations of Chiral $N=2$ $D=10$ Supergravity,"
Nucl. Phys. {\bf B226} (1983) 269.

\bibitem{n2def}
A.~Buchel,
``Compactifications of the $N = 2^*$ flow,''
arXiv:hep-th/0302107.

\bibitem{wip} Work in progress.

\end{thebibliography}
\end{document}